\documentclass[12pt]{iopart}

\usepackage{amssymb}

\usepackage{graphicx}

\usepackage{pstricks}

\begin{document}

\title{Novel surface universality classes with strong anisotropy}
\author{B. Schmittmann$^{1}$, Gunnar Pruessner$^{1,2}$, and H.K. Janssen$^{3}$}
\date{\today}

\begin{abstract}
Using renormalized field theory, we examine the dynamics
of a growing surface, driven by an obliquely incident particle beam.
Its projection on the reference (substrate) plane selects a 
``parallel'' direction, so that the evolution equation 
for the surface height becomes anisotropic. The phase diagram 
of the model is controlled by the properties of an effective anisotropic 
surface tension. Our renormalization group analysis suggests the
existence of a line of continuous transitions and a line of (potentially)
first-order transitions, which meet at a multicritical point.
The full scaling behavior for the continuous line and the 
multicritical point is discussed in detail. Two novel universality 
classes for scale-invariant surface fluctuations are found. 
\end{abstract}

			
\address{$^1$Center for Stochastic Processes in Science and Engineering, \\
Department of Physics, Virginia Tech, Blacksburg, VA 24061-0435, USA;\\
$^{2}$Department of Physics (CMTH), Imperial College, London SW7 2BW, UK;\\
$^{3}$Institut f\"{u}r Theoretische Physik III, Heinrich-Heine-Universit\"{a}t,
40225 D\"{u}sseldorf, Germany}
\ead{\mailto{schmittm@vt.edu}, \mailto{gunnar.pruessner@physics.org},
\mailto{janssen@thphy.uni-duesseldorf.de}}

\submitto{\JPA}



\section{Introduction}

The capability to grow nanoscale layers of atoms or molecules on a substrate
forms an important part of modern nanotechnology. Particles emitted from a
source are deposited on the growing surface whose characteristics are
determined by parameters such as growth temperature, surface diffusion, bulk
relaxation, and particle attachment vs detachment rates. In addition to 
being clearly relevant for device applications, such surface growth 
problems also constitute an important class of generic nonequilibrium 
phenomena \cite{HHZ:1995,Krug:1997} which cannot be described in terms of 
standard equilibrium ensemble theory.

A key goal of both theoretical and experimental investigations is the
characterization and understanding of the resulting surface morphology,
especially in terms of its statistical properties such as spatial and temporal
height-height correlations. While local features may play an important role
for device performance, large-scale, long-time characteristics offer better
opportunities for theoretical understanding. They are often universal, i.e.,
independent of microscopic details, depending only on global constraints such
as symmetries and conservation laws. Here, we will explore the consequences
of a \emph{reduced} symmetry, which results if the particles arrive under an 
oblique (non-normal) angle.

Let us briefly review different classes of models for surface growth. The
first issue is whether surface overhangs and shadowing effects can be
neglected. For models with near-grazing incidence, this is usually a poor
approximation. Focusing on vapor-deposited thin films, Meakin and Krug
\cite{Meakin:1988,MeakinKrug:1990,MeakinKrug:1992} considered ballistic
deposition under these conditions. They found that columnar patterns form
which shield parts of the growing surface from incoming particles and
characterized these structures in terms of anisotropic scaling exponents,
differentiating parallel and transverse directions
\cite{MeakinKrug:1992,KrugMeakin:1989}. In contrast, for normal or near-normal
incidence, overhangs and shadowing can often be neglected, and a description
in term of a stochastic equation of motion for a single-valued height variable
becomes possible. A second important issue is whether particles, once attached
to the surface, can desorb again. If desorption does occur, the surface
relaxation is not conserved, and the \emph{leading} nonlinearity in the
Langevin equation is of the Kardar-Parisi-Zhang form
\cite{KardarParisiZhang:1986}. Models of this type are most appropriate for
chemical vapor deposition (see, e.g., \cite{Bales:1989,Ojeda:2000}) 
and sputtering processes \cite{Makeev:2002}. Anisotropies have also 
been included by different authors 
\cite{Wolf:1991,Hwa:1992,TF:2002}. In
contrast, desorption can often be neglected in MBE-type processes which occur
at low growth temperatures (see, e.g., \cite{Lagally:1990}). In this case, the
surface relaxation terms in the Langevin equation conserve the number of
particles. For models of this type, the universal behavior depends on whether
the incident beam is normal to the substrate or not. For normal incidence, the
system remains isotropic on large scales while oblique incidence generates
anisotropies which persist under coarse-graining. While the former case has
been investigated extensively (see, e.g., \cite{dasSarma:1996,Krug:2002}), 
only partial results are available for the latter \cite{MarsiliETAL:1996}. 
Noting that the reduced symmetry generates
novel leading nonlinearities suggesting potentially distinct universal
behavior, we consider surface growth under oblique incidence more
systematically in the following.

Our analysis begins with a model first suggested by Marsili \textit{et al.}
\cite{MarsiliETAL:1996} to describe MBE-type or ballistic deposition processes
with oblique particle incidence. Based on a coarse-grained description in
terms of a Langevin equation, the model adopts an idealized perspective
\cite{iso-MBE,Villain:1991,Janssen:1996}, neglecting
particle desorption and bulk defect formation. 
As a result, the surface relaxation becomes mass-conserving and can be written
as the gradient of a current, in an appropriate reference
frame. In contrast, the noise term in the Langevin equation is not
mass-conserving, since it models the random arrivals of particles.
Anisotropies emerge naturally in the Langevin equation, since the particle
beam selects a preferred (``parallel'') direction in the substrate plane.
Extending the analysis of \cite{MarsiliETAL:1996}, we explore the consequences
of an effective anisotropic surface tension. Generated by the interplay of
interatomic interactions and kinetic effects, such as Schwoebel barriers, it
plays a central role for the long-time, long-distance properties of the model.
Due to the anisotropy, it will generically take different values, denoted
$\tau_{\Vert}$ and $\tau_{\bot}$, in the parallel and transverse directions.
Depending on which of these parameters vanishes first, ripple-like surface
instabilities are expected, aligned transverse to the soft direction. This
leads to four different regimes with potentially scale-invariant behaviors. We
analyze these four regimes, identify the scale-invariant ones, and discuss the
associated anisotropic roughness exponents, using techniques from renormalized
field theory. The original theory of \cite{MarsiliETAL:1996} is recovered only
if both couplings, $\tau_{\Vert}$ \textit{and} $\tau_{\bot}$, vanish
simultaneously. 

The paper is organized as follows. We first present the underlying Langevin
equation for a single-valued height field and briefly review the physical
origin of its constituents. We also give a full discussion of its symmetries
which are important for the subsequent analysis. We then turn to the
renormalization group (RG) analysis. Identifying a set of effective coupling
constants, and exploiting a Ward identity following from a (continuous) tilt
invariance, we compute the naively divergent one-particle irreducible
(1PI)\ vertex functions to one-loop order near the upper critical dimension.
We analyze the four different cases separately:\ (o) both $\tau_{\Vert}$ and
$\tau_{\bot}$ are positive; (i) $\tau_{\Vert}$ remains positive while
$\tau_{\bot}$ vanishes; (ii) $\tau_{\Vert}$ vanishes while $\tau_{\bot}$
remains positive; and finally, (iii) both $\tau_{\Vert}$ and $\tau_{\bot}$
vanish simultaneously. A renormalization group equation allows us to derive
the scaling properties of correlation and response functions for those cases
which possess infrared (IR)\ stable fixed points. We identify a complete set
of critical exponents, including \emph{four different} roughness exponents 
for each case.
Two of these characterize real-space scans along and transverse to the beam
direction, and the remaining two are needed to describe scattering (i.e.
momentum space) data with parallel or transverse momentum transfer.


\section{The model}

If surface overhangs and shadowing effects are neglected, the
surface can be described by a single-valued height field, $h(\mathbf{r},t)$.
Here, $\mathbf{r}$ denotes a $d$-dimensional vector in a reference (substrate)
plane, and $t$ denotes time. $h$ is measured
along the $z$-axis which is normal to the substrate plane. The no-overhang
assumption can be justified \textit{a posteriori} provided the interface
roughness exponents are less than unity \cite{MarsiliETAL:1996}. The time
evolution of the interface is described by a Langevin equation, combining
deterministic terms, $G[h]$, and the effects of randomness, $\eta$, in the
form
\begin{equation}
\partial_{t}h=G[h]+\eta\label{generic_Langevin}%
\end{equation}
The conserved nature of the (deterministic) surface relaxation can be captured
by writing $G[h]=\nabla\mathbf{\cdot F}[h]$. Several different terms, arising
from surface diffusion and the incident flux, contribute to $G[h]$ and are
discussed in the following.

The incident particle current has a normal component $J_{z}$ and a component
parallel to the substrate plane, $\mathbf{J}_{\Vert}$. For particles of finite
size $r_{o}$, the effective flux must be measured at a distance $r_{o}$ normal
to the surface \cite{finite}.
Since we focus on the long time, large distance characteristics of the growing
surface, the corresponding contributions to $G[h]$ can be written as a
gradient expansion, neglecting higher order terms:
\begin{equation}
G_{\mathrm{flux}}[h]=-J_{z}+\mathbf{J}_{\parallel}\cdot\mathbf{\nabla}%
h+r_{o}J_{z}\nabla^{2}h-r_{o}\mathbf{J}_{\parallel}\cdot\left(  \mathbf{\nabla
}h\right)  \,\nabla^{2}h \label{music}%
\end{equation}
The first two terms of Eq.~(\ref{music}) can be removed by transforming 
into an appropriate co-moving frame, 
via $h(\mathbf{r},t)\rightarrow h(\mathbf{r+J}_{\parallel}t,t)-J_{z}t$. 
Next, we consider the effects of diffusion \textit{along} the surface. This
restriction generates a \emph{quartic} term, of the form \cite{Mullins:1963}
\begin{equation}
G_{\mathrm{diff}}\ =-\mu\nabla^{4}h
\end{equation}
Several comments are in order. First, we note the presence of the term
$r_{o}J_{z}\nabla^{2}h\equiv$ $\tau\nabla^{2}h$ in Eq. (\ref{music}). Similar
terms can arise from a step edge (Schwoebel) barrier
\cite{SchwoebelShipsey:1966,Schwoebel:1969,Villain:1991}, coming with a
negative sign, or a surface tension with a positive sign. This allows for the
possibility that the net contribution $\tau\nabla^{2}h$ might vanish, playing
the role of a critical parameter. Even if we follow \cite{MarsiliETAL:1996}
and set it to zero in the bare theory, it is actually generated under RG
transformations and therefore intrinsically present. Second, the full
rotational symmetry within the $d$-dimensional space of the substrate is
broken by the nonlinear term in Eq.~(\ref{music}). As a consequence,
any coarse-graining of the microscopic 
theory should give rise to different couplings in the parallel and transverse
subspaces, and this is in fact borne out under the renormalization group. 
If the full anisotropy is incorporated into the theory, restricting
rotational invariance to the $(d-1)$-dimensional \emph{transverse} subspace,
the Langevin equation (\ref{generic_Langevin}) takes the form
\begin{eqnarray}
\gamma^{-1}\partial_{t}h &=& \tau_{\parallel}\partial_{\parallel}^{2}h+\tau
_{\perp}\nabla_{\perp}^{2}h-\mu_{\parallel}\partial_{\parallel}^{4}%
h-2\mu_{\times}\nabla_{\perp}^{2}\partial_{\parallel}^{2}h-\mu_{\perp}%
(\nabla_{\perp}^{2})^{2}h \nonumber \\
&+& \left(  \partial_{\parallel}h\,\right)  \bigl(
\lambda_{\parallel}\partial_{\parallel}^{2}h+\lambda_{\perp}\nabla_{\perp}%
^{2}h\bigr)  +\eta\label{full_problem}%
\end{eqnarray}
after some minor renamings. Here, $\partial_{\parallel}$ ($\nabla_{\bot})$
denotes derivatives (gradients) in the parallel (transverse) subspaces. It
should be noted that the nonlinearity has also been split into two distinct
terms, with couplings $\lambda_{\parallel}$ and $\lambda_{\perp}$. The
Langevin noise $\eta(\mathbf{r},t)$ models the random particle aggregation at
the surface, with zero average and second moment
\begin{equation}
\left\langle \eta(\mathbf{r},t)\eta(\mathbf{r}^{\prime},t^{\prime
})\right\rangle =2\gamma^{-1}\delta(\mathbf{r}-\mathbf{r}^{\prime}%
)\ \delta(t-t^{\prime})\ . \label{noise_correl}%
\end{equation}
In principle, one could also incorporate a noise term of the form
$-2(D_{\Vert}\partial_{\parallel}^{2}+D_{\bot}\nabla_{\perp}^{2}%
)\delta(\mathbf{r}-\mathbf{r}^{\prime})\ \delta(t-t^{\prime})$ generated by
(conserved) particle diffusion. In comparison to the aggregation noise, this
contribution turns out to be irrelevant. However, one could imagine situations
where diffusion is extremely fast, and aggregation occurs only very
infrequently, so that Eq.~(\ref{noise_correl}) should be replaced by its
diffusive counterpart. This ``conserved''\ ideal MBE
\cite{SunETAL:1989,Janssen:1996} is not considered here, and will be left for
future study.

Eq. (\ref{full_problem}) can be written in terms of surface currents
$j_{\parallel}$ and $\mathbf{j}_{\perp}$, according to
\begin{equation}
\gamma^{-1}\partial_{t}h=-\partial_{\parallel}j_{\parallel}-\nabla_{\perp
}\mathbf{j}_{\perp}+\eta \label{currents}%
\end{equation}
with
\begin{eqnarray}
j_{\parallel}  &  =-\partial_{\parallel}\bigl(  \tau_{\parallel}%
h-\mu_{\parallel}\partial_{\parallel}^{2}h-\mu_{\times}\nabla_{\perp}%
^{2}h\bigr)  -\frac{\lambda_{\parallel}}{2}\bigl(  \partial_{\parallel
}h\bigr)  ^{2}+\frac{\lambda_{\perp}}{2}\bigl(  \nabla_{\perp}h\bigr)
^{2}\,, \nonumber \\
\mathbf{j}_{\perp}  &  =-\nabla_{\perp}\bigl(  \tau_{\perp}h-\mu_{\perp}%
\nabla_{\perp}^{2}h-\mu_{\times}\partial_{\parallel}^{2}h\bigr)
-\lambda_{\perp}\bigl(  \nabla_{\perp}h\bigr)  \,\bigl(  \partial_{\parallel
}h\bigr)  \,. \nonumber
\end{eqnarray}

\begin{figure}[ptb]
\begin{center}
\includegraphics[width=7.5cm]{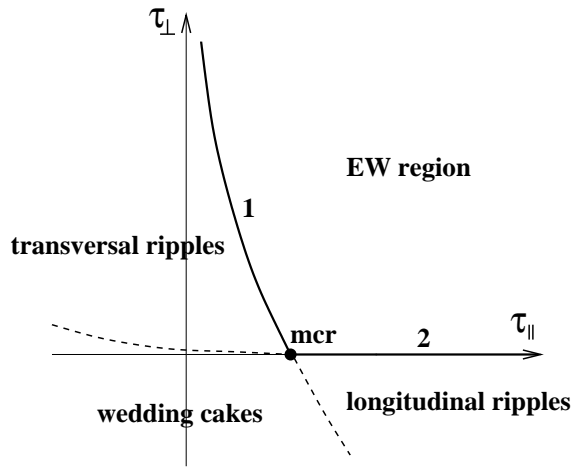}  
\end{center}
\caption{ Sketch of the phase diagram in $(\tau_{\parallel},\tau_{\perp})$ 
space. EW denotes the Edwards--Wilkinson region. 
The line labelled $1$ marks a line of (possibly) discontinuous transitions, 
and line $2$ corresponds to continuous
transitions, both to inhomogeneous structures. 
The dashed lines denote very qualitative transition linesn between 
different structures, and mcr labels the multicritical point. }
\label{PhDia}%
\end{figure}

Eq.~(\ref{full_problem}), or its equivalent form, Eq. (\ref{currents}), is the
starting point for the field-theoretic analysis. Its behavior is determined by
the leading terms in the gradient expansion. We can gain some qualitative
insight by considering the linear terms first. To ensure stability of the
homogeneous phase, the coupling constants $\mu_{\parallel}$ and $\mu_{\perp}$
must be positive, and the cross-coupling $\mu_{\times}$ may not exceed
$-(\mu_{\parallel}\mu_{\perp})^{1/2}$ . The two couplings $\tau_{\parallel}$
and $\tau_{\perp}$ play the role of critical parameters. If both are positive,
the nonlinearities become irrelevant, and the problem reduces to the
(Gaussian)\ Edwards-Wilkinson equation \cite{EdwardsWilkinson:1982}. In
contrast, if one or both of them vanish, the surface develops characteristic
spatial structures. These take the form of ripples (reminiscent of corrugated
roofing), if only one of the two couplings goes through zero; if both
couplings become negative, the surface develops mounds or ``wedding
cakes''\ \cite{Krug:1997}. Focusing only on the onset of these instabilities,
four different cases emerge which are discussed systematically in the
following section: 
(o) the linear theory $\tau_{\parallel}>0$, $\tau_{\perp}>0$; 
(i) a line of continuous transitions $\tau_{\parallel}>0$, 
$\tau_{\perp}\rightarrow0$; 
(ii) a line of \textit{possibly first} order transitions 
$\tau_{\parallel}\rightarrow0$, $\tau_{\perp}>0$; 
and (iii) the multicritical (critical end-) point 
$\tau_{\parallel}\rightarrow0$, $\tau_{\perp}\rightarrow0$. 
The qualitative phase diagram is shown in
Fig.~(\ref{PhDia}).

\section{Renormalization group analysis}

\subsection{The dynamic functional}

To unify the discussion, we first recast the Langevin equation
(\ref{currents}) as a dynamic field theory. In this formulation, all
statistical averages are expressed as path integrals, with weight
$\exp(-\mathcal{J})$. In order to treat correlation and response functions in
an analogous manner, it is convenient to introduce a response field $\tilde
{h}(\mathbf{r},t)$, and formulate the dynamic response functional, following
standard methods \cite{Janssen:1976,DeDominicis:1976,Janssen:1992}:
\begin{equation}
\mathcal{J}[\tilde{h},h]=\gamma\int d^{d}x \, dt \Big\{  \tilde{h}\Big[
\gamma^{-1}\partial_{t}h+\partial_{\parallel}j_{\parallel}+\nabla_{\perp
}\mathbf{j}_{\perp}\Big]  -\tilde{h}^{2}\Big\}  \,. \label{def_J}%
\end{equation}

Before we turn to any computational details, we argue that the model 
is renormalizable. In other words, in each loop order we need to 
add only those primitive counterterms which can be absorbed in a 
renormalization of the fields and the coupling constants.
As a first step in this procedure, we consider the symmetries exhibited by
(\ref{def_J}) since they place constraints on possible counterterms. The
invariance against a shift of coordinates in the $z$-direction leads to the
symmetry $h(\mathbf{r},t)\rightarrow h(\mathbf{r},t)+a$. Hence, the surface
height $h$ arises in the functional only in form of derivatives. Second, the
functional is invariant under tilts of the surface by an infinitesimal
``angle''\ $\mathbf{b}$, i.e., $h(\mathbf{r},t)\rightarrow h(\mathbf{r}%
,t)+\mathbf{b}\cdot\mathbf{r}$ provided the tilt is accompanied by an
appropriate change of the parameters, namely, $\tau_{\parallel}\rightarrow
\tau_{\parallel}^{\prime}=\tau_{\parallel}-b_{\Vert}\lambda_{\parallel}$ and
$\tau_{\perp}\rightarrow\tau_{\perp}^{\prime}=\tau_{\perp}-b_{\Vert}%
\lambda_{\perp}$. Further, particle conservation on the surface lead to the
invariance transformation $\tilde{h}(\mathbf{r},t)\rightarrow\tilde
{h}(\mathbf{r},t)+c$, $h(\mathbf{r},t)\rightarrow h(\mathbf{r},t)+2c\gamma t$
. Therefore, we have a $(d+2)$-dimensional (infinitesimal) symmetry group
\begin{eqnarray}
h(\mathbf{r},t)  &  \rightarrow h(\mathbf{r},t)+a+\mathbf{b}\cdot
\mathbf{r}+2c\gamma t\,,\label{h-sym} \qquad
\tilde{h}(\mathbf{r},t)  &  \rightarrow\tilde{h}(\mathbf{r},t)+c\,,
\nonumber
\end{eqnarray}
which does not change the form of the dynamic functional or the functional
integration measure, provided it is accompanied by an appropriate
transformation of the parameters $\tau_{\parallel}$ and $\tau_{\perp}$. In
addition, we have a discrete symmetry, namely, inversion invariance in the
$(x_{\Vert},z)$ subspace:%
\begin{eqnarray}
h(x_{\Vert},\mathbf{r}_{\bot},t)\rightarrow-h(-x_{\Vert},\mathbf{r}_{\bot
},t)\,,\qquad\tilde{h}(x_{\Vert},\mathbf{r}_{\bot},t)\rightarrow-\tilde
{h}(-x_{\Vert},\mathbf{r}_{\bot},t)\,. \nonumber
\end{eqnarray}
Finally, of course, we have isotropy in the transverse subspace. These
symmetries can be considered as the fundamental defining elements of ideal MBE processes.

The renormalizability of our model can now be argued, following standard
methods \cite{Amit:1984,Zinn-Justin:1996}. In each loop order, the successive
construction of the perturbation series produces only \emph{primitive}
ultraviolet (UV)\ divergences, provided that all counterterms constructed in
lower orders are included to cancel \emph{non-primitive} divergences which
appear in subdiagrams. If the regularization procedure respects the symmetries
of the model, then the remaining primitive divergences also preserve the
symmetries. Thus, these divergences have the same form as the relevant terms
of the initial model and their counterterms can be absorbed in a
renormalization of the initial parameters. This presumes, of course, that all
relevant interaction terms (i.e., composite fields with coupling constants of
positive naive dimension) with the correct symmetries have been included in
the initial dynamic functional, and all irrelevant couplings have been
omitted. With regards to our model, a straightforward analysis of the relevant
diagrams shows that we have indeed captured all relevant terms in the original
Langevin equation (\ref{full_problem}). Hence, the model is renormalizable.

\subsection{Elements of perturbation theory: the four different cases}

In this section, we assemble the basic components of the perturbative
analysis, leaving technical details to the Appendix. We first write the
dynamic functional, $\mathcal{J}[\tilde{h},h]$ as the sum of two parts.
$\mathcal{J}_{0}[\tilde{h},h]$ contains the Gaussian terms of the theory,
\begin{eqnarray}
\mathcal{J}_{0}[\tilde{h},h] &=& 
\gamma\int d^{d}x \, dt \Big\{  \tilde
{h}\bigl(  \gamma^{-1}\partial_{t}+\mu_{\parallel}\partial_{\parallel}^{4}%
+\mu_{\perp}(\nabla_{\perp}^{2})^{2} \nonumber \\
&+&2\mu_{\times}\partial_{\parallel}%
^{2}\nabla_{\bot}^{2}-\tau_{\parallel}\partial_{\parallel}^{2}-\tau_{\perp
}\nabla_{\bot}^{2}\bigr)  h-\tilde{h}^{2}\Big\}  \label{def_J_0}%
\end{eqnarray}
while $\mathcal{J}_{1}[\tilde{h},h]$ takes the nonlinear interactions into
account:\
\begin{equation}
\mathcal{J}_{1}[\tilde{h},h]=-\gamma\int d^{d}x \, dt\ \tilde{h}%
\,(\partial_{\parallel}h)\,\bigl(  \lambda_{\parallel}\partial_{\parallel}%
^{2}h+\lambda_{\perp}\nabla_{\perp}^{2}h\bigr)  \label{def_J_int}%
\end{equation}
Arbitrary correlation and response functions can now be computed as functional
averages with statistical weight $\mathcal{J}[\tilde{h},h]$:
\begin{eqnarray}
\langle\circ\rangle=\int\mathcal{D}[h]\mathcal{D}[\tilde{h}]\circ
\mathrm{e}^{-\mathcal{J}[\tilde{h},h]}\,. \nonumber
\end{eqnarray}

Due to the anisotropy, it is possible to rescale parallel and transverse
lengths independently. Considering a simple rescaling of \textit{parallel}
lengths first, via $x_{\parallel}\rightarrow\alpha x_{\parallel}$, the
functional remains invariant provided $h\rightarrow\alpha^{-1/2}h$, $\tilde
{h}\rightarrow\alpha^{-1/2}\tilde{h}$ and $\mu_{\parallel}\rightarrow
\alpha^{4}\mu_{\parallel}$, $\tau_{\parallel}\rightarrow\alpha^{2}%
\tau_{\parallel}$, $\mu_{\times}\rightarrow\alpha^{2}\mu_{\times}$ while
$\lambda_{\parallel}\rightarrow\alpha^{7/2}\lambda_{\parallel}$ and
$\lambda_{\perp}\rightarrow\alpha^{3/2}\lambda_{\perp}$. Likewise, if only
\textit{transverse}\textrm{\ }lengths are rescaled, via $\mathbf{r}_{\perp
}\rightarrow\beta\mathbf{r}_{\perp}$, the functional remains invariant
provided $h\rightarrow\beta^{-(d-1)/2}h$, $\tilde{h}\rightarrow\beta
^{-(d-1)/2}\tilde{h}$ and $\mu_{\perp}\rightarrow\beta^{4}\mu_{\perp}$,
$\tau_{\perp}\rightarrow\beta^{2}\tau_{\perp}$, $\mu_{\times}\rightarrow
\beta^{2}\mu_{\times}$ while $\lambda_{\parallel}\rightarrow\beta
^{(d-1)/2}\lambda_{\parallel}$ and $\lambda_{\perp}\rightarrow\beta
^{(d+3)/2}\lambda_{\perp}$. In addition, there is an overall length scale
$\kappa^{-1}$ which accounts for the dimension of fields and coupling
constants under the renormalization group. When we consider the three
different cases below, these scale invariances will allow us to define
appropriate effective expansion parameters, in order to eliminate redundant
couplings \cite{Wegner:1976}.

Our goal is to compute the Green functions $G_{N,\tilde{N}}(\{\mathbf{r}%
,t\})=\langle\lbrack h]^{N}[\tilde{h}]^{\tilde{N}}\rangle^{\mathrm{cum}}$,
i.e., the cumulant averages of all possible products of fields $h(\mathbf{r}%
,t)$, $\tilde{h}(\mathbf{r},t)$, or equivalently, the associated one-particle
irreducible (1PI) vertex functions, $\Gamma_{\tilde{N},N}(\{\mathbf{r},t\})$,
with $\tilde{N}$ ($N$) $h$ ($\tilde{h}$-) amputated
legs, order by order in a diagrammatic perturbation expansion. 
The notation $\{\mathbf{r},t\}$ is short-hand for the full
space- and time-dependence of these functions.
As a first step in
this process, we collect the elements of perturbation theory and introduce
their graphic representation. Using $\langle\cdot\rangle_{0}$ to denote
averages taken with the (Gaussian) weight $\mathcal{J}_{0}$, we will need 
the bare propagator,
$G_{0}(\mathbf{q},t)\,\delta(\mathbf{q}-\mathbf{q}^{\prime})\equiv\langle
h(\mathbf{q},t)\tilde{h}(-\mathbf{q}^{\prime},0)\rangle_{0} \,,$
as well as the bare correlator,
$C_{0}(\mathbf{q},t)\,\delta(\mathbf{q}-\mathbf{q}^{\prime})\equiv\langle
h(\mathbf{q},t)h(-\mathbf{q}^{\prime},0)\rangle_{0} \,,$
both in spatial Fourier space. In the following, we write $\int_{\vec
{q},\omega}\dots$ as short-hand for $\int\frac{d^{d}q}{(2\pi)^{d}}%
\int\frac{d\omega}{2\pi}\dots$. Defining
\begin{equation}
\Gamma(\mathbf{q})\equiv\mu_{\parallel}q_{\parallel}^{4}+\mu_{\perp
}(\mathbf{q}_{\bot}^{2})^{2}+2\mu_{\times}q_{\parallel}^{2}\mathbf{q}_{\bot
}^{2}+\tau_{\parallel}q_{\parallel}^{2}+\tau_{\perp}\mathbf{q}_{\bot}^{2}
\label{full_momentum_part}%
\end{equation}
we find%
\begin{eqnarray}
G_{0}(\mathbf{q},t) &  =\theta(t)\,\exp\bigl( -\Gamma(\mathbf{q})\gamma
t\bigr) \,,\nonumber\\
C_{0}(\mathbf{q},t) &  =\Gamma(\mathbf{q})^{-1}G_{0}(\mathbf{q}%
,|\,t|)\,.\label{def_bare_props}%
\end{eqnarray}
Here the Heaviside function $\theta(t)$ is defined with $\theta(0)=0$.
Diagrammatically, these two functions are represented as lines:
\label{diagram_propagators}%
\begin{eqnarray}
G_{0}(\mathbf{q},t) &  =\begin{pspicture}(-0.2,0)(2.2,1)\rput(-0.1,0.05){$t$}
\rput(0,0){\psline{->}(2,0)(1,0)\psline(0,0)(2,0)\rput(1,0.3){$\mathbf{q}
$}\rput(2.15,0.05){$0$}} \end{pspicture} \nonumber \\
C_{0}(\mathbf{q},t) &  =\begin{pspicture}(-0.2,0)(2.2,1)\rput(-0.1,0.05){$t$}
\rput(0,0){\psline(2,0)(1,0)\psline(0,0)(2,0)\rput(1,0.3){$\mathbf{q}%
$}\rput(2.15,0.05){$0$}} \end{pspicture} \nonumber
\end{eqnarray}
Turning to the interaction terms, it is convenient to rewrite them in a
symmetrized form. In Fourier space, the expression for the three-point vertex
reads
\begin{eqnarray}
V(\mathbf{q}_{1},\mathbf{q}_{2},\mathbf{q}_{3}) &\equiv & i\gamma\Big\{
\lambda_{\parallel}\,q_{1\parallel}\,q_{2\parallel}\,
q_{3\parallel}+ \nonumber \\
&+& \lambda_{\perp}\Big[ \mathbf{q}_{1\bot}\cdot\bigl( q_{2\parallel}%
\,\mathbf{q}_{3\bot}+\mathbf{q}_{2\bot}\,q_{3\parallel}\bigr) -q_{1\parallel
}\,\mathbf{q}_{2\bot}\cdot\mathbf{q}_{3\bot}\Big] \Big\}
\,\label{full_vertices}%
\end{eqnarray}
with $\mathbf{q}_{1}+\mathbf{q}_{2}+\mathbf{q}_{3}=0$. Our sign convention is
such that all momenta attached to a vertex are incoming. Diagrammatically, the
whole vertex is represented as
\begin{eqnarray}
\begin{pspicture}(-2.2,-1.0)(4.5,1.0) \rput(-2.2,0){ $V(\mathbf{q}%
_{1}, \mathbf{q}_{2}, \mathbf{q}_{3})=$} \rput(-0.5,-0){\rput(0,0){\psline{<-}%
(0,0)(1,0)\rput(0.72,0.3){ \rput(0,0){$\mathbf{q}_1$}}}}
\rput(0.5,0){\rput{45}%
(0,-0){\rput(0,0){\psline(0,0)(1,0) \rput(1,0.4){$\mathbf{q}_3$}}}}
\rput(0.5,0){\rput{-45}%
(0,-0){\rput(0,0){\psline(0,0)(1,0) \rput(1,-0.4){$\mathbf{q}_2$}}}}
\end{pspicture} \nonumber 
\end{eqnarray}
Returning to Eqs.~(\ref{def_J_0}) and (\ref{def_J_int}), we now perform
dimensional analysis. In the (infrared) limit of small momenta and
frequencies, the Gaussian part of the dynamic functional is dominated by
different terms, depending on the behavior of the control parameters
$\tau_{\parallel}$ and $\tau_{\perp}$. 

\noindent {\it Case} (o). 
If both $\tau_{\parallel}$ and $\tau_{\perp}$ are finite and
positive, the theory turns out to be purely Gaussian. Quartic derivatives can
be neglected in the infrared limit; as a result, $\Gamma(\mathbf{q})$
simplifies to $\Gamma^{(1)}(\mathbf{q})\equiv\tau_{\parallel}q_{\parallel}%
^{2}+\tau_{\perp}\mathbf{q}_{\bot}^{2}$. Since both parallel and transverse
momenta appear only quadratically here, it is natural to choose a momentum
scale $\kappa$ such that $q_{\parallel}\propto\left|  \mathbf{q}_{\bot
}\right|  \propto\kappa$, and $\tau_{\parallel}\propto\tau_{\perp}%
\propto\kappa^{0}$. Time $\gamma t$ scales as $\kappa^{-2}$, and the fields
have dimensions $h(\mathbf{r},t)\propto\kappa^{(d-2)/2}$ and $\tilde
{h}(\mathbf{r},t)\propto\kappa^{(d+2)/2}$. The nonlinear couplings scale as
$\lambda_{\parallel}\propto\lambda_{\perp}\propto\kappa^{-d/2}$ and are
therefore irrelevant in any dimension $d>0$. The resulting theory is a simple
anisotropic generalization of the Edwards-Wilkinson equation
\cite{EdwardsWilkinson:1982},
\begin{equation}
\gamma^{-1}\partial_{t}h=\tau_{\parallel}\partial_{\parallel}^{2}h+\tau
_{\perp}\nabla_{\perp}^{2}h+\eta\ .
\end{equation}
The anisotropies in the quadratic terms affect only nonuniversal amplitudes
and can be removed by a simple rescaling, without losing any information of
interest. As is well known, the two-point correlation function scales as
\begin{equation}
C(\mathbf{r},t)=\left|  \mathbf{r}\right|  ^{2-d}c(t/\left|  \mathbf{r}%
\right|  ^{2})
\end{equation}
from which one immediately reads off the (isotropic) roughness exponent
$\chi=(2-d)/2$ and the dynamic exponent $z=2$. For a detailed discussion see
\cite{Krug:1997}. Since this case is so familiar, we will not consider 
it any further.

\noindent {\it Case} (i). 
If $\tau_{\parallel}$ remains finite and positive but
$\tau_{\perp}$ is infinitesimal, the two leading terms in the dynamic
functional are $\mu_{\perp}\tilde{h}(\nabla_{\perp}^{2})^{2}h$ and
$\tau_{\parallel}\tilde{h}\partial_{\parallel}^{2}h$. Hence, even in the
Gaussian theory, parallel and transverse momenta already scale 
differently, reflected in the choice $\left|  \mathbf{q}_{\bot}\right|
\propto\kappa$ and $q_{\parallel}\propto\kappa^{2}$. If we introduce an
anisotropic scaling exponent $\Delta$ via $q_{\parallel}\propto\kappa
^{1+\Delta}$, we recognize that $\Delta=1$ at the tree level. Continuing with
the two leading terms, we note $\tau_{\parallel}\propto\mu_{\perp}%
\propto\kappa^{0}$, and set $\mu_{\perp}=1$ via a transverse rescaling with an
appropriate $\beta$. Time scales as $\kappa^{-4}$, and $\mu_{\parallel}$
$\propto\mu_{\times}\propto\kappa^{-2}$ are both irrelevant. The only strongly
relevant parameter of the theory is $\tau_{\perp}\propto\kappa^{2}$.
Introducing the effective dimension $D \equiv d+1$, one finds $h(\mathbf{r}%
,t)\propto\kappa^{(D-4)/2}$ and $\tilde{h}(\mathbf{r},t)\propto\kappa
^{(D+4)/2}$. For the nonlinear couplings, we obtain $\lambda_{\perp}%
\propto\kappa^{(4-D)/2}$ and $\lambda_{\parallel}\propto\kappa^{-D/2}$. Since
$D$ is clearly positive, the coupling $\lambda_{\parallel}$ becomes
irrelevant. The upper critical dimension $d_{c}$ for the theory is determined
by $\lambda_{\perp}$, via $0=4-D$ which leads to $d_{c}=3$. The invariant
dimensionless effective expansion parameter is $\tau_{\parallel}^{-3/4}%
\lambda_{\perp}/\kappa^{(3-d)/2}$ as shown by the rescaling $\lambda_{\perp
}\rightarrow\alpha^{3/2}\lambda_{\perp}$, $\tau_{\parallel}\rightarrow
\alpha^{2}\tau_{\parallel}$. The resulting propagators are controlled by
$\Gamma(\mathbf{q})\equiv\tau_{\perp}\mathbf{q}_{\bot}^{2}+\tau_{\parallel
}q_{\parallel}^{2}+(\mathbf{q}_{\bot}^{2})^{2}$ where the $(\mathbf{q}_{\bot
}^{2})^{2}$-term determines the UV behavior and $\tau_{\bot}$ plays the role
of the IR cutoff. At the tree level, this case corresponds to a critical line
parameterized by $\tau_{\parallel}$.

\noindent {\it Case} (ii). 
Here, $\tau_{\perp}$ remains finite while $\tau_{\parallel}$
vanishes. The Gaussian part of the functional is dominated by $\tilde
{h}\partial_{\parallel}^{4}h$ and $\tau_{\perp}\tilde{h}\nabla_{\perp}^{2}h$.
Again, parallel and transverse momenta scale differently, but their roles are
now reversed: $q_{\parallel}\propto\kappa$ and $\left|  \mathbf{q}_{\bot
}\right|  \propto\kappa^{1+\Delta}$, with $\Delta=1$ at the tree level. Both
$\mu_{\parallel}\ $and $\tau_{\perp}$ scale as $\kappa^{0}$, while $\mu
_{\perp}$ and $\mu_{\times}$ are irrelevant. Defining the effective dimension
as $D\equiv2(d-1)+1$, one still has $h(\mathbf{r},t)\propto\kappa^{(D-4)/2}$
and $\tilde{h}(\mathbf{r},t)\propto\kappa^{(D+4)/2}$; we also recover $\gamma
t\propto\kappa^{-4}$. Now, however, the strongly relevant perturbation is
$\tau_{\parallel}\propto\kappa^{2}$. The two nonlinearities switch roles so
that $\lambda_{\parallel}\propto\kappa^{(6-D)/2}$ and $\lambda_{\perp}%
\propto\kappa^{-D/2}$. Clearly, $\lambda_{\perp}$ is irrelevant in all
dimensions while $\lambda_{\parallel}$ becomes marginal at the upper critical
dimension $d_{c}=7/2$. The invariant dimensionless effective expansion
parameter follows from the rescalings as $\mu_{\parallel}^{-7/8}\tau_{\bot
}^{-(d-1)/4}\lambda_{\Vert}/\kappa^{(7-2d)/2}$. The momentum dependence of the
propagators simplifies to $\Gamma(\mathbf{q})\equiv\tau_{\parallel
}q_{\parallel}^{2}+\tau_{\perp}\mathbf{q}_{\bot}^{2}+\mu_{\parallel
}q_{\parallel}^{4}$. As in the previous case, at the tree level this situation
corresponds to a critical line parametrized by $\tau_{\perp}$. However, we
will see below that the order of the transition may well become first order
once fluctuations are included.

\noindent {\it Case} (iii). 
Finally, we consider the multicritical point where both
$\tau_{\perp}$ and $\tau_{\parallel}$ vanish. Both momenta scale identically,
as $q_{\parallel}\propto\left|  \mathbf{q}_{\bot}\right|  \propto\kappa$, so
that $\Delta=0$ at the tree level. Again, we may set $\mu_{\perp}=1$. One
obtains $\gamma t\propto$ $\kappa^{-4}$, $\tau_{\parallel}\propto\tau_{\perp
}\propto\kappa^{2}$, and $h(\mathbf{r},t)\propto\kappa^{(D-4)/2}$, $\tilde
{h}(\mathbf{r},t)\propto\kappa^{(D+4)/2}$, with $D=d$. The full propagators
come into play. Both nonlinear couplings, $\lambda_{\parallel}$ and
$\lambda_{\perp}$, have the same upper critical dimension $d_{c}=6$. The
effective expansion parameters are $w\equiv\mu_{\times}/\sqrt{\mu_{\parallel}%
}$, $u_{\parallel}\equiv\mu_{\parallel}^{-7/8}\lambda_{\parallel}%
/\kappa^{(6-d)/2}$, and\ $u_{\perp}\equiv\mu_{\parallel}^{-3/8}\lambda_{\perp
}/\kappa^{(6-d)/2}$. In this case, the anisotropy exponent $\Delta$ vanishes
at the tree level.

In the following, we analyze the three nontrivial cases in a
one-loop approximation, using dimensional regularization combined with minimal
subtraction \cite{Amit:1984,Zinn-Justin:1996}. The essential components 
are the 1PI vertex functions $\Gamma_{\tilde{N},N}%
(\{\mathbf{q},\omega\})$. Focusing on the ultraviolet singularities, 
only those $\Gamma_{\tilde{N},N}$
with positive engineering dimension are to be considered. Taking into account
the symmetries and the momentum-dependence carried by the derivatives on the
external legs, the set of naively divergent vertex functions is reduced to
$\Gamma_{1,1}$ and $\Gamma_{1,2}$. Specifically, at the upper critical
dimension of cases (ii) and (iii), $\Gamma_{1,1}$ is quadratically divergent
and $\Gamma_{1,2}$ is marginal. In case (i), $\Gamma_{1,1}$ is already
marginal after accounting for the external momenta.


\subsection{Perturbation theory.}

In this section, we summarize the calculation of $\Gamma_{1,1}$ and
$\Gamma_{1,2}$, to one loop order. Details are relegated to the Appendix.

The main features which recur in all three cases are the following. If
$\Gamma_{1,1}$ is quadratically divergent, this divergence is first removed by
an additive renormalization, i.e., a shift of the critical parameter(s).
Specifically, the true critical point is located through the singularity of
the parallel and/or transverse static susceptibilities, $\chi_{\Vert}$ and
$\chi_{\bot}$, defined by
\begin{eqnarray}
\gamma\mu_{\parallel}\,\chi_{\Vert}^{-1}&\equiv& \lim_{q_{\parallel}\rightarrow
0}q_{\parallel}^{-2}\Gamma_{1,1}(q_{\parallel},\mathbf{q}_{\bot}%
=\mathbf{0},\omega=0) \nonumber \\
\gamma\chi_{\bot}^{-1} &\equiv& \lim
_{\mathbf{q}_{\bot}\rightarrow0}(\mathbf{q}_{\bot}^{2})^{-1}\Gamma
_{1,1}(q_{\parallel}=0,\mathbf{q}_{\bot},\omega=0) \label{susc}%
\end{eqnarray}
We find that nontrivial shifts are only required for cases (ii) and (iii),
and only for $\tau_{\parallel}$.

The remaining logarithmic divergences are then computed using dimensional
regularization, so that ultraviolet divergences appear as simple poles in
$\varepsilon\equiv d_{c}-d$. In a minimal subtraction scheme, 
we focus exclusively on these poles and their amplitudes.
Since the nonlinearities are cubic in the field, the expansion is organized in
powers of $u_{\parallel}^{2}$, $u_{\perp}^{2}$, and $u_{\parallel}u_{\perp}$;
i.e., the first correction to the tree level is always \textit{quadratic} for
$\Gamma_{1,1}$ and \textit{cubic} for $\Gamma_{1,2}$. The tilt invariance
leads to a Ward identity connecting $\Gamma_{1,1}$ and $\Gamma_{1,2}$,
namely,
\label{ward_identity}%
\begin{eqnarray}
-i\left.  \frac{\partial}{\partial q_{\Vert}^{\prime}}\right|  _{\mathbf{q}%
^{\prime},\omega^{\prime}=0}\Gamma_{1,2}\left(  \mathbf{q},\omega
;\mathbf{q}^{\prime},\omega^{\prime}\right)   &  =\lambda_{\Vert
}\frac{\partial\Gamma_{1,1}(\mathbf{q},\omega)}{\partial\tau_{\parallel}%
}+\lambda_{\perp}\frac{\partial\Gamma_{1,1}(\mathbf{q},\omega)}{\partial
\tau_{\perp}}\\
-i\left.  \frac{\partial}{\partial\mathbf{q}_{\bot}^{\prime}}\right|
_{\mathbf{q}^{\prime},\omega^{\prime}=0}\Gamma_{1,2}\left(  \mathbf{q}%
,\omega;\mathbf{q}^{\prime},\omega^{\prime}\right)   &  =0 \nonumber
\end{eqnarray}
Here, $(\mathbf{q},\omega)$ and $(\mathbf{q}^{\ \prime},\omega^{\prime})$
denote the momenta and frequencies labelled with subscripts $2$ and $3$ in
Eq.~(\ref{full_vertices}).
 By virtue of this symmetry, it will be sufficient
to compute $\Gamma_{1,1}$ in order to extract the renormalization of
\textit{all parameters}, including those of the coupling constants
$\lambda_{\Vert}$ and $\lambda_{\perp}$. More specifically, the tilt
transformation $h(\mathbf{r},t)\rightarrow h(\mathbf{r},t)+\mathbf{b}%
\cdot\mathbf{r}$, $\tau_{\parallel}\rightarrow\tau_{\parallel}-b_{\Vert
}\lambda_{\parallel}$, $\tau_{\perp}\rightarrow\tau_{\perp}-b_{\Vert}%
\lambda_{\perp}$ shows that the parameter $\mathbf{b}$ renormalizes as the
field $h$ itself. Hence, the term $\lambda_{\parallel}\,h$ renormalizes
exactly like $\tau_{\parallel}$, and $\lambda_{\perp}h$ like $\tau_{\perp}$.

Next, we collect a few general properties of the diagrams contributing to
$\Gamma_{1,1}(\mathbf{q},\omega)$ which follow from the special form of the
interactions, Eq.~(\ref{def_J_int}). We first consider diagrams where the
external $h$-leg carries two transverse derivatives. Such diagrams can arise
only from the interaction $\lambda_{\perp}\tilde{h}\,(\partial_{\parallel
}h)\,\nabla_{\perp}^{2}h$. Thanks to particle conservation, all these graphs
sum in such a way as to generate at least one spatial derivative on the
external $\tilde{h}$-leg. Moreover, either parallel inversion or transverse
rotation symmetry guarantees that, in fact, there are at least two such
derivatives. Thus, there are no corrections to the term $\gamma\tau_{\bot
}\tilde{h}\nabla_{\bot}^{2}h$ in $\mathcal{J}_{0}$, Eq.~(\ref{def_J_0}).
Similarly, since external $\tilde{h}$- and $h$-legs always come with at least
one spatial derivative, there are no perturbative contributions to the terms
$\tilde{h}\partial_{t}h$ and $\gamma\tilde{h}^{2}$. Hence, these three terms
are not renormalized. Further, if the coupling constant $\lambda_{\perp}$ is
zero, as in case (ii), only diagrams with parallel derivatives on all legs are
generated, and so operators carrying no parallel, or mixed parallel and
transverse, derivatives suffer no renormalization. In contrast, when
$\lambda_{\perp}\neq0$, there is no such simplification since this vertex can
contribute to purely ``parallel'' terms. If both coupling constants
$\lambda_{\Vert}$ and $\lambda_{\perp}$ are nonzero, they mix under the RG,
and we have to renormalize the couplings by matrices, rather than simple
multiplicative factors.

With these considerations in mind, we propose the following renormalization
scheme:
\begin{eqnarray}
h\rightarrow\mathring{h}&=& Z^{1/2}h \qquad \quad \,\,\,\,
\tilde{h}\rightarrow\mathring{\tilde{h}} = Z^{-1/2}\tilde{h}
\qquad \quad \,\,\,\,
\gamma\rightarrow\mathring{\gamma} = Z\gamma \nonumber \\
\tau_{\perp}\rightarrow\mathring{\tau}_{\perp} &=& Z^{-1}\tau_{\perp} 
\qquad \quad \,\,
\tau_{\parallel}\rightarrow\mathring{\tau}_{\parallel} = Z^{-1}\bigl(
Z_{\tau}\tau_{\parallel}+\mu_{\parallel}^{1/2}Y\tau_{\perp}\bigr)
+\mathring{\tau}_{\parallel,c} \nonumber \\
\mu_{\times}\rightarrow\mathring{\mu}_{\times} &=& Z^{-1}Z_{\times}\mu_{\times}
\qquad \mu_{\Vert}\rightarrow\mathring{\mu}_{\Vert} 
= Z^{-1}Z_{\mu}\mu_{\Vert} \label{gen_ren_scheme}  
\end{eqnarray}
$\mathring{\tau}_{\parallel,c}$ accounts for the shift of the critical point
in cases (ii) and (iii). The Ward identities then lead to the
renormalizations
\begin{eqnarray}%
\lambda_{\perp}\rightarrow\mathring{\lambda}_{\perp} & =Z^{-3/2}\lambda
_{\perp} \qquad \qquad \lambda_{\Vert}\rightarrow\mathring{\lambda
}_{\Vert} & =Z^{-3/2}\bigl(  Z_{\tau}\lambda_{\Vert}+Y\lambda_{\perp}\bigr)
\,. \label{lambda_ren}%
\end{eqnarray}
The five renormalization factors $Z$, $Z_{\tau}$, $Z_{\times}$, $Z_{\mu}$, and
$Y$ must be calculated from the $\varepsilon$-poles of the dimensionally
regularized perturbation series of $\Gamma_{1,1}(\mathbf{q},\omega).$ In
minimal subtraction, they are of the form $Z_{\circ}=1+\sum_{n=1}^{\infty
}Z_{\circ}^{(n)}/\varepsilon^{n}$, $Y=\sum_{n=1}^{\infty}Y^{(n)}%
/\varepsilon^{n}$, where the $Z_{\circ}^{(n)}$ and $Y^{(n)}$ are expansions
in the renormalized effective coupling constants. Here, the subscript 
$\circ$ stands as a placeholder for any member of the set
of indices. The individual
characteristics of the three cases of interest are now treated sequentially.
\newline
\\
\noindent {\it Case} (i): $\tau_{\perp}\rightarrow0$ and $\tau_{\parallel}%
\neq0$. 
This is the simplest non-trivial case. Only one parameter, $\tau_{\perp}$,
needs to be tuned to access criticality. Since $\lambda_{\Vert}$ is
irrelevant, it may be set to zero. Neglecting all other irrelevant terms as
well, the functional simplifies to
\begin{equation}
\mathcal{J}=\gamma\int d^{d}x \, dt \Big\{  \tilde{h}\bigl[
\gamma^{-1}\partial_{t}+(\nabla_{\perp}^{2})^{2}-\tau_{\parallel}%
\partial_{\parallel}^{2}-\tau_{\perp}\nabla_{\perp}^{2}\bigr]  h-\tilde{h}%
^{2}-\lambda_{\perp}\tilde{h}(\partial_{\parallel}h)\nabla_{\perp}^{2}h\Big\}
\end{equation}
In this case $\mu_{\Vert}=\mu_{\times}=\lambda_{\Vert}=Y=0$. All divergences
are parameterized by $\varepsilon\equiv d_{c}-d$ where $d_{c}=3$.

Following standard methods (for details see \ref{ParFinitePerpVanish}%
), we obtain the following expression for the singular part of the two-point
vertex function,
\begin{equation}
\Gamma_{1,1}(\mathbf{q},\omega)=i\omega+\gamma\big[
\tau_{\parallel}q_{\parallel}^{2}+(\mathbf{q}_{\bot}^{2})^{2}+\tau_{\perp
}\mathbf{q}_{\bot}^{2}\big] +\gamma\frac{u^{2}}{8\varepsilon}\big[
2\tau_{\parallel}q_{\parallel}^{2}-(\mathbf{q}_{\bot}^{2})^{2}\big]
+O(u^{4}) \label{Gamma11_i}%
\end{equation}
where
\begin{equation}
u\equiv A_{\varepsilon}\tau_{\parallel}^{-3/4}\lambda_{\perp}\kappa
^{-\varepsilon/2}\label{def_u_i}%
\end{equation}
is the effective expansion parameter. $A_{\varepsilon}$ is a geometric
factor which appears in all integrals: 
\[
A_{\varepsilon}^{2}\equiv\frac{S_{d-1}}{(2\pi)^{d}}\sqrt{\pi}\Gamma\left(
\frac{1-\varepsilon}{2}\right)  \Gamma\left(  \frac{1+\varepsilon}{2}\right)
\]
and $S_{d}$ is the surface area of the $d$-dimensional unit sphere.

The renormalized vertex function $\Gamma_{1,1}$ is defined by demanding that
\begin{eqnarray}
\Gamma_{1,1}(\mathbf{q},\omega,\gamma,\tau_{\perp},\tau_{\parallel}%
,u,\kappa)\equiv\mathring{\Gamma}_{1,1}(\mathbf{q},\omega,\mathring{\gamma
},\mathring{\tau}_{\perp},\mathring{\tau}_{\parallel},\mathring{\lambda
}_{\perp}) \nonumber 
\end{eqnarray}
be pole-free. One finds
\begin{eqnarray}
Z  &  =1+\frac{u^{2}}{8\varepsilon}+O(u^{4})\,, \qquad 
Z_{\tau}  &  =1-\frac{u^{2}}{4\varepsilon}+O(u^{4})\,.\label{res_Z_i}
\end{eqnarray}
The corresponding Wilson functions are defined as the logarithmic derivatives
of the associated $Z$-factors, at constant bare quantities, i.e.,
$  \zeta_{\circ}\equiv\kappa\partial_{\kappa}\ln Z_{\circ}|_{\mathrm{bare}}\,$.
The logarithmic derivatives of the control parameters and of the
relaxation coefficient are then given by
\label{def_zetas}%
\begin{eqnarray}
&  \kappa\partial_{\kappa}\ln\tau_{\Vert}|_{\mathrm{bare}}=\zeta-\zeta_{\tau
}\,,\nonumber \\
&  \kappa\partial_{\kappa}\ln\tau_{\bot}|_{\mathrm{bare}}=\zeta\,,\\
&  \kappa\partial_{\kappa}\ln\gamma|_{\mathrm{bare}}=-\zeta\,, \nonumber
\end{eqnarray}
The flow of the dimensionless effective coupling constant $u$ under
renormalization is controlled by the Gell-Mann--Low function,
\begin{equation}
\beta\equiv\kappa\partial_{\kappa}u|_{\mathrm{bare}}=
u \big[  -\frac{\varepsilon
}{2}+\frac{3}{4}\left(  \zeta+\zeta_{\tau}\right)  \big]  \,.
\label{beta_u_i}%
\end{equation}
The renormalizability of the theory demands that all the Wilson functions be
free of $\varepsilon$-poles; moreover, in minimal subtraction it can be shown
\cite{Amit:1984} that the perturbative corrections do not contain $\varepsilon$
any more. It is easy to demonstrate that all Wilson
functions are determined by the first term, $Z_{\circ}^{(1)}$, in the
Laurent expansion of the $Z$-factors. So, we obtain
\begin{eqnarray}
\zeta(u)  &  =-\frac{u}{2}\frac{\partial Z^{(1)}}{\partial u}=-\frac{u^{2}}%
{8}+O(u^{4})\,, \nonumber \\
\zeta_{\tau}(u)  &  =-\frac{u}{2}\frac{\partial Z_{\tau}^{(1)}}{\partial
u}=\frac{u^{2}}{4}+O(u^{4})\,,\nonumber \\
\beta(u)  &  = u\big[  -\frac{\varepsilon}{2}+\frac{3}{32}u^{2}+O(u^{4})\big]
\,. \label{beta_i}%
\end{eqnarray}

The renormalization group equation (RGE) for the Green functions simply states
that the bare theory is independent of the external momentum scale $\kappa$:
\begin{eqnarray}
0&=&\kappa\frac{d}{d\kappa}\mathring{G}_{N,\tilde{N}}(\{\mathbf{r}%
,t\};\mathring{\tau}_{\bot},\mathring{\tau}_{\Vert},\mathring{\gamma
},\mathring{\lambda}_{\perp}) \nonumber \\
&=&\kappa\frac{d}{d\kappa}Z^{(N-\tilde{N})/2}%
G_{N,\tilde{N}}(\{\mathbf{r},\omega\};\tau_{\bot},\tau_{\Vert};u,\gamma
,\kappa)\,.
\end{eqnarray}
Explicitly, this partial differential equation becomes
\begin{equation}
\Big[  \kappa\frac{\partial}{\partial\kappa}+\beta\frac{\partial}{\partial
u}-\zeta\gamma\frac{\partial}{\partial\gamma}+\zeta\tau_{\perp}\frac{\partial
}{\partial\tau_{\perp}}+(\zeta-\zeta_{\tau})\tau_{\parallel}\frac{\partial
}{\partial\tau_{\parallel}}+\frac{\zeta}{2}(N-\tilde{N})\Big]  G_{N,\tilde{N}}%
=0\,. \label{RGE_i}%
\end{equation}
It will lead to asymptotic scaling, provided $\beta(u)$ possesses an infrared
stable fixed point $u^{\ast}$, i.e., a solution of $\beta(u^{\ast})=0$ with
$\beta^{\prime}(u^{\ast})>0$. To one loop order, Eq.~(\ref{beta_i}) allows for
a single fixed point with the desired properties, namely,
\begin{equation}
u^{\ast}=\pm4\sqrt{\frac{\varepsilon}{3}}\,\bigl(  1+O(\varepsilon)\bigr)  \ ,
\label{eq:fps_case_tperp0}%
\end{equation}
where the sign is determined by the sign of the initial coupling constant
$\lambda_{\perp}$. Clearly, this $u^{\ast}$ is non-zero in the $\varepsilon
$-expansion. Under the (reasonable) assumption that the full (resummed) series
for $u^{\ast}(\varepsilon)$ is non-zero at the integer values of $\varepsilon$
corresponding to physical dimensions, Eq.~(\ref{beta_u_i}) gives us the exact
relation
\begin{equation}
\zeta(u^{\ast})+\zeta_{\tau}(u^{\ast})=\frac{2\varepsilon}{3} \label{exact_i}
\end{equation}
In the following, Wilson $\zeta$-functions, evaluated at the fixed point, will
simply be denoted by a superscript $\ast$.

The solution of the RGE (\ref{RGE_i}) is easily found by applying the method
of characteristics. At the stable fixed point $u^{\ast}$ we obtain the scaling
solution
\begin{eqnarray}
G_{N,\tilde{N}}(\{\mathbf{r},t\};\tau_{\bot},\tau_{\Vert},
\gamma,\kappa)=l^{\zeta^{\ast}(N-\tilde{N})/2}G_{N,\tilde{N}}(\{\mathbf{r}%
,t\};l^{\zeta^{\ast}}\tau_{\bot},l^{\zeta^{\ast}-\zeta_{\tau}^{\ast}}
\tau_{\Vert},l^{-\zeta^{\ast}}\gamma,l\kappa) 
\nonumber
\end{eqnarray}
where $l$ is an arbitrary flow parameter. The parallel and transverse scale
invariances, together with dimensional analysis, give us an additional scaling
relation for the Green functions, namely
\begin{eqnarray}
G_{N,\tilde{N}}(\{\mathbf{r},t\};&\tau_{\perp}&,\tau_{\parallel},u_{\perp
},\gamma,\kappa)   =\alpha^{(N+\tilde{N})/2}\kappa
^{N(D-4)/2+\tilde{N}(D+4)/2} \nonumber \\
& \times & G_{N,\tilde{N}}(\{\alpha\kappa^{2}x_{\parallel}%
,\kappa\mathbf{r}_{\perp},\gamma\kappa^{4}t\};\tau_{\perp}/\kappa^{2}%
,\alpha^{2}\tau_{\parallel},u_{\perp};1,1)  \nonumber
\end{eqnarray}
Combining these two relations, choosing $\alpha
=l^{(\zeta_{\tau}^{\ast}-\zeta^{\ast})/2}$ and recalling $D=d+1$, we arrive at
the final scaling form for the Green functions. Suppressing unneeded
arguments, we can choose to write them in the form 
\begin{eqnarray}
G_{N,\tilde{N}}(\{\mathbf{r},t\};\tau_{\bot})  = 
l^{\delta_{N,\tilde{N}}}G_{N,\tilde{N}}(\{l^{1+\Delta}x_{\parallel
},l\mathbf{r}_{\perp},l^{z}t\};l^{-1/\nu}\tau_{\bot})\, \label{G_i-exp}%
\end{eqnarray}
where the scaling exponents $\Delta$ and $\eta$ can be
expressed in terms of Wilson functions, evaluated at the fixed point:
\begin{eqnarray}
\Delta &  =1+\frac{\zeta_{\tau}^{\ast}-\zeta^{\ast}}{2}\,,\qquad\eta
= \zeta^{\ast} \label{Delta-eta_i}
\end{eqnarray}
Since we have only two independent renormalizations, the remaining 
exponents follow from scaling laws, namely,
\begin{eqnarray}
z =4-\eta\,,\qquad\qquad1/ \nu=2-\eta\,,\label{Res-exp_i}
\end{eqnarray}
and 
\begin{equation}
\delta_{N,\tilde{N}}=\frac{N}{2}(d+\Delta -4 +\eta)
+\frac{\tilde{N}}{2}(d+\Delta +4-\eta)  \label{deltaNN}
\end{equation}
Eq. (\ref{exact_i}), provided it holds, can be exploited to give 
another exponent identity: 
\begin{eqnarray}
\Delta+\eta &  =2-d/3\,.\label{more_scaling}
\end{eqnarray}
To conclude, only a \textit{single} exponent, e.g., $\eta$ has to be computed
order by order in perturbation theory. Then, all others follow from exponent
identities which are exact, at least within perturbation theory. Our one-loop
calculation results in
\begin{equation}
\eta=-2\varepsilon/3+O(\varepsilon^{2})\,. \label{eta_i}%
\end{equation}

\vspace{0.5cm}
\noindent {\it Case} (ii): $\tau_{\parallel}\rightarrow0$ and $\tau_{\perp}%
\neq0$. This is the second non-trivial case. Neglecting irrelevant
terms, the dynamic response functional simplifies to
\begin{equation}
\mathcal{J}[\tilde{h},h]=\gamma\int d^{d}x\, dt\ \Big\{  \tilde{h}\Big[
\gamma^{-1}\partial_{t}+\mu_{\parallel}\partial_{\parallel}^{4}-\tau
_{\parallel}\partial_{\parallel}^{2}-\tau_{\perp}\nabla_{\perp}^{2}\Big]
h-\tilde{h}^{2}-\lambda_{\parallel}\tilde{h}(\partial_{\parallel}%
h)\partial_{\parallel}^{2}h\Big\} 
\end{equation}
In this case, we define $\varepsilon\equiv d_{c}-d$ with $d_{c}=7/2$.

Since $\Gamma_{1,1}(\mathbf{q},\omega)$ is quadratically divergent in this
case, an additive renormalization of the critical parameter, $\tau_{\parallel
}$, is required (see \ref{ParVanishPerpFinite}).
Once this is accounted for, we obtain the following
one-loop expression for the singular part of $\Gamma_{1,1}$:
\begin{equation}
\Gamma_{1,1}(\mathbf{q},\omega)_{\mathrm{pole}}=i\omega+\gamma\Big(
\mu_{\parallel}q_{\parallel}^{4}+\tau_{\Vert}q_{\parallel}^{2}
+\tau_{\bot}\mathbf{q}%
_{\bot}^{2}\Big)  +\gamma\frac{u^{2}}{2\varepsilon}\Big(  \mu_{\parallel
}q_{\parallel}^{4}+\tau_{\parallel}q_{\parallel}^{2}\Big)  + \dots
\label{eq:Gamma11_ii}%
\end{equation}
where $u$ is the effective expansion parameter now defined as
\begin{equation}
u\equiv B_{\varepsilon}\mu_{\parallel}^{-7/8}\tau_{\perp}^{-(d-1)/4}%
\lambda_{\parallel}\kappa^{-(7-2d)/2} \label{effective_coupl_case_tpar0}%
\end{equation}
with a geometric factor
\[
B_{\varepsilon}^{2}\equiv\frac{S_{d-1}}{2(2\pi)^{d}}\Gamma\left(
\frac{d-1}{2}\right)  \Gamma\left(  \frac{d}{2}\right)  \Gamma(1+\varepsilon
)\ .
\]
Eq.~(\ref{eq:Gamma11_ii}) gives rise to renormalizations of
$\mu_{\parallel}$ and $\tau_{\parallel}$. Since the vertex $\lambda
_{\parallel}$ carries only parallel momenta in this case, there are no
contributions of order $\mathbf{q}_{\bot}^{2}$ at any order, implying the
absence of field renormalizations. Hence, $Z=1$. The remaining poles are
absorbed into the renormalization factors
\begin{eqnarray}
Z_{\mu}  &  =1-\frac{u^{2}}{2\varepsilon}+O(u^{4})\,,\nonumber\\
Z_{\tau}  &  =1-\frac{u^{2}}{2\varepsilon}+O(u^{4})\,. \label{Z_ii}%
\end{eqnarray}
Defining the Gell-Mann--Low function of $u$, as well as the Wilson functions,
in analogy to the previous case, we obtain
\begin{eqnarray}
\zeta_{\mu}  &  =\kappa\partial_{\kappa}\ln Z_{\mu}|_{\mathrm{bare}%
}=u^{2}+O(u^{4})\,,\nonumber\\
\zeta_{\tau}  &  =\kappa\partial_{\kappa}\ln Z_{\tau}|_{\mathrm{bare}%
}=u^{2}+O(u^{4}%
)\,,\label{zeta_ii}\\
\beta &  \equiv l\partial_{l}u_{\parallel}|_{\mathrm{bare}}=u\big[
-\varepsilon+  \frac{7}{8}\zeta_{\mu}-\zeta_{\tau}  \big]
\,=u\big[  -\varepsilon-\frac{1}{8}u^{2}+O(u^{4})\big]  \nonumber
\end{eqnarray}
The fixed point equation $\beta(u^{\ast})=0$ does not yield a stable real
fixed point, at least to this order in perturbation theory. It remains an open
question whether such a fixed point, and the associated scaling properties,
might emerge at higher orders. Restricting ourselves to our current results,
the absence of an infrared stable fixed point might suggest a first order
transition. A more detailed analysis of the underlying mean-field theory or a
careful computational study would be required to shed more light on this
issue. Both are beyond the scope of this article.
\newline

\noindent {\it Case} (iii): $\tau_{\parallel}\rightarrow0$ and $\tau_{\perp}%
\rightarrow0$. Finally, we turn to the analysis of the 
multicritical point. This situation was previously considered in
\cite{MarsiliETAL:1996}, using a momentum shell decimation scheme. This
procedure requires considerable care for field theories with strong anisotropy
and nonlinearities carrying multiple derivatives, since the corrections depend
on the way in which the hard momentum cutoff is implemented. Moreover, the
hard momentum shell cutoff introduces long-ranged correlations on the scale of
the cutoff momentum which must be handled very carefully. Whatever the source
of the discrepancies, even after meticulous checks we were unable to reproduce the
earlier results.

The full functional, Eqs.~(\ref{def_J_0}) and (\ref{def_J_int}), as well as
the full (bare) propagators and correlators, Eq.~(\ref{def_bare_props}), now
come into play. To ensure the stability of the critical theory at the tree
level, we demand $\mu_{\parallel} \, q_{\parallel}^{4}+2\mu_{\times}
q_{\parallel}^{2} \, \mathbf{q}_{\bot}^{2}+(\mathbf{q}_{\bot}^{2})^{2}\geq0$.
This limits the physical range of $\mu_{\parallel}$ and $\mu_{\times}$ to
$\mu_{\parallel}>0$ and $\mu_{\times}>-\sqrt{\mu_{\parallel}}$. To complicate
matters further, both nonlinear couplings, $\lambda_{\parallel}$ and
$\lambda_{\perp}$, are marginal at the upper critical dimension $d_{c}=6$.
However we know that, thanks to the Ward identity, all renormalizations can
still be obtained from the two-point function, $\Gamma_{1,1}$.

A first analysis of diagrams contributing to $\Gamma_{1,1}(\mathbf{q},\omega)$
shows that two of them are quadratically divergent. Since both of them carry a
momentum prefactor of $q_{\Vert}^{2}$, they can be absorbed in an additive
renormalization, i.e.., a shift of the bare control parameter $\mathring{\tau
}_{\parallel}$, as remarked before. No shift of the bare $\mathring{\tau
}_{\bot}$ is needed.

Leaving details to \ref{ParVanishPerpVanish}, our one-loop result for
$\Gamma_{1,1}$ is now considerably more complex, thanks to the presence of
both vertices. In order to eliminate redundant parameters, it is again
convenient to define invariant dimensionless coupling constants, guided by the
parallel and transverse rescalings:
\label{effective_couplings_caseIII}%
\begin{eqnarray}
u_{\Vert}   \equiv C_{\varepsilon}\mu_{\parallel}^{-7/8}(1+w)^{-5/4}%
\lambda_{\Vert}\kappa^{-(6-d)/2}\,, \qquad   w \equiv\frac{\mu_{\times}}%
{\mu_{\parallel}^{1/2}}\,,\nonumber \\
u_{\bot}   \equiv C_{\varepsilon}\mu_{\parallel}^{-3/8}(1+w)^{-5/4}%
\lambda_{\bot}\kappa^{-(6-d)/2}\,, 
\end{eqnarray}
where the constant $C_{\varepsilon}$ is defined by 
\[
C_{\varepsilon}^{2}\equiv\frac{S_{d-1}}{(2\pi)^{d}}\frac{\pi\Gamma
(1+\varepsilon/2)}{32\sqrt{2}}%
\]
The definition of the effective coupling constants contains a suitable
$w$-dependent denominator, which is common to all integrals. Collecting the
results from \ref{ParVanishPerpVanish}, we find the singular parts
\begin{eqnarray}
\Gamma_{1,1}(\mathbf{q},\omega)_{\mathrm{pole}} &=&  i\omega+\gamma\Big[
q_{\parallel}^{2}(\mu_{\parallel}q_{\parallel}^{2}+\tau_{\Vert})+\mathbf{q}%
_{\bot}^{2}(\mathbf{q}_{\bot}^{2}+\tau_{\bot})+2\mu_{\times}q_{\parallel}%
^{2}\mathbf{q}_{\bot}^{2}\Big] \nonumber\\
&-&\gamma\mu_{\parallel}^{1/2}q_{\parallel}^{2}\biggl[  u_{\Vert}^{2}\Big(
\frac{B_{1}}{\varepsilon}\mu_{\parallel}^{-1/2}\tau_{\parallel}+\frac{B_{1}%
}{\varepsilon}\tau_{\perp}\Big)  +u_{\bot}^{2}\Big(  \frac{B_{2}}{\varepsilon
}\mu_{\parallel}^{-1/2}\tau_{\parallel}+\frac{B_{3}}{\varepsilon}\tau_{\perp
}\Big)  \biggr] \nonumber\\
&-& \gamma\mu_{\parallel}q_{\parallel}^{4}\Big[  u_{\Vert}^{2}\frac{C_{1}%
}{\varepsilon}+u_{\bot}^{2}\frac{C_{2}}{\varepsilon}\Big]  -\gamma
(\mathbf{q}_{\bot}^{2})^{2}u_{\bot}^{2}\frac{A_{1}}{\varepsilon}%
\label{Gamma11_iii}\\
&-& \gamma\mu_{\parallel}^{1/2}q_{\parallel}^{2}\mathbf{q}_{\bot}^{2}\Big[
u_{\Vert}^{2}\frac{D_{1}}{\varepsilon}+u_{\bot}^{2}\frac{D_{2}}{\varepsilon
}+u_{\Vert}u_{\bot}\frac{D_{3}}{\varepsilon}\Big]  \,.\nonumber
\end{eqnarray}

Here, $A_{1}$, $B_{1}$, ..., $D_{3}$ are $w$-dependent functions whose
explicit forms are listed in \ref{ParVanishPerpVanish}. 

According to the general renormalization scheme, Eq. (\ref{gen_ren_scheme}),
the full set of five renormalization factors now comes into play. At one loop
order, we read off
\label{Zfactors_caseIII_a}%
\begin{eqnarray}
Z  &  =1+A_{1}\frac{u_{\bot}^{2}}{\varepsilon}+O(u^{4})\,,\nonumber\\
Z_{\tau}  &  =1+B_{1}\frac{u_{\Vert}^{2}}{\varepsilon}+B_{2}\frac{u_{\bot}%
^{2}}{\varepsilon}+O(u^{4})\,,\nonumber\\
Y  &  =B_{1}\frac{u_{\Vert}^{2}}{\varepsilon}+B_{3}\frac{u_{\bot}^{2}%
}{\varepsilon}+O(u^{4})\,,\nonumber\\
Z_{\mu}  &  =1+C_{1}\frac{u_{\Vert}^{2}}{\varepsilon}+C_{2}\frac{u_{\bot}^{2}%
}{\varepsilon}+O(u^{4})\,,\label{Z_iii}\\
Z_{\times}  &  =1+D_{1}\frac{u_{\Vert}^{2}}{2w\varepsilon}+D_{2}\frac{u_{\bot
}^{2}}{2w\varepsilon}+D_{3}\frac{u_{\Vert}u_{\bot}}{2w\varepsilon}%
+O(u^{4})\,.\nonumber
\end{eqnarray}
Here, $O(u^{4})$ is short-hand for the two-loop corrections which are of
fourth order in the couplings $u_{\Vert}$ and $u_{\bot}$. Defining
$\zeta_{\circ}\equiv\kappa\partial_{\kappa}\ln Z_{\circ}|_{\mathrm{bare}%
}=-\frac{1}{2}(u_{\Vert}\partial_{u_{\Vert}}+u_{\perp}\partial_{u_{\perp}%
})Z_{\circ}^{(1)}$, $\zeta_{y}=-\frac{1}{2}(u_{\Vert}\partial_{u_{\Vert}%
}+u_{\perp}\partial_{u_{\perp}})Y^{(1)}$ as before, one obtains to one loop
order
\begin{eqnarray}
\zeta &  =-2(1+w)u_{\bot}^{2}\,,\qquad\zeta_{\tau}=3u_{\parallel}%
^{2}-(5+2w)u_{\bot}^{2}\,,\nonumber\\
\zeta_{y}  &  =3u_{\parallel}^{2}-3(7+10w+4w^{2})u_{\bot}^{2}\,,\nonumber\\
\zeta_{\mu}  &  =(3+w)u_{\parallel}^{2}-(5+9w+10w^{2}+4w^{3})u_{\bot}%
^{2}\,,\label{zeta_iii}\\
\zeta_{\times}  &  =\frac{(7+5w)}{10w}u_{\parallel}^{2}-\frac{(49+65w+26w^{2}%
)}{10w}u_{\bot}^{2}-\frac{(10+2w)}{10w}u_{\parallel}u_{\perp}\,.\nonumber
\end{eqnarray}

We now consider the Gell-Mann--Low functions for the three dimensionless 
effective couplings $u_{\Vert}$, $u_{\bot}$, and $w$:
\label{beta_iiia}
\begin{eqnarray}
\beta_{\parallel} & \equiv & \left.  \kappa\partial_{\kappa}u_{\Vert}\right|
_{\mathrm{bare}}=\Big(  -\frac{\varepsilon}{2}+\frac{5}{8}\zeta-\zeta_{\tau
}+\frac{7}{8}\zeta_{\mu}-\frac{5\beta_{w}}{4(1+w)}\Big)  u_{\Vert}-\zeta
_{y}u_{\bot}  \nonumber\\
\beta_{\perp}    &\equiv& \left.  \kappa\partial_{\kappa}u_{\bot}\right|
_{\mathrm{bare}}=\Big(  -\frac{\varepsilon}{2}+\frac{9}{8}\zeta+\frac{3}{8}%
\zeta_{\mu}-\frac{5\beta_{w}}{4(1+w)}\Big)  u_{\bot}  \label{beta_iii}\\
\beta_{w}    &\equiv&  \kappa\partial_{\kappa}w|_{\mathrm{bare}}
=\Big(  \frac{1}{2}%
\zeta-\zeta_{\times}+\frac{1}{2}\zeta_{\mu}\Big)  w   \nonumber
\end{eqnarray}
We seek a set of fixed points $(w^{\ast},u_{\Vert}^{\ast},u_{\bot}^{\ast})$
such that all three $\beta$-functions vanish simultaneously. The eigenvalues
of the linearized system, evaluated in the vicinity of each fixed point,
provide information about its stability. If all of them are positive, the
fixed point is infrared stable; otherwise, it is unstable in one 
or more directions. 

There is only one nontrivial, physical (i.e., real) infrared stable fixed
point given by
\begin{equation}
w^{\ast}=2\sqrt{\frac{3}{5}}-1+O(\varepsilon)\,, \,\, u_{\Vert}^{\ast}%
=\pm\sqrt{\frac{7\sqrt{15}+25}{11}\,}\varepsilon^{1/2}+O(\varepsilon
^{3/2})\,, \,\, u_{\bot}^{\ast}=0 \label{fp_iii}%
\end{equation}
For $u_{\bot}^{\ast}=0$, we find another root $w^{\ast}=-1-\frac{2}{5}%
\sqrt{15}<-1$, but this solution must be excluded since it leads to a linearly
unstable theory (see the discussion after Eq.~(\ref{currents})). There are
numerous fixed points with $u_{\bot}^{\ast}\neq0$ and $u_{\Vert}^{\ast}\neq0$,
but these are all infrared unstable or nonphysical, in the sense that either
$u_{\Vert}^{\ast}$ or $u_{\bot}^{\ast}$ are imaginary or $w^{\ast}<-1$. In
contrast to earlier results by \cite{MarsiliETAL:1996}, we do not find a
physically viable fixed point with $u_{\bot}^{\ast}\neq0$. We have also
carefully considered the possibility of ``degenerate fixed points'' with
$w^{\ast}=0$ or $w^{\ast}=\infty$; none of these are stable.

Returning briefly to Eq. (\ref{Gamma11_iii}), we note that the critical
parameter $\tau_{\parallel}$ appears only in the combination $\tilde{\tau
}_{\parallel}\equiv\mu_{\parallel}^{-1/2}\tau_{\parallel}$, which is actually
the appropriate scale-invariant form. This suggests a slight modification
of the general renormalization scheme, Eq.~(\ref{gen_ren_scheme}), so that the
renormalizations of the critical control parameters are now written as
\begin{equation}
\mathring{\tilde{\tau}}_{\parallel}=Z^{-1/2}Z_{\mu}^{-1/2}\bigl(  Z_{\tau
}\tilde{\tau}_{\parallel}+Y\tau_{\perp}\bigr)  \,,\qquad\mathring{\tau}%
_{\perp}=Z^{-1}\tau_{\perp}\,, \label{Z_tau}%
\end{equation}
Their derivatives can be expressed in terms of the $\zeta$-functions of Eq.
(\ref{zeta_iii}):\
\begin{eqnarray}
\left.  \kappa\partial_{\kappa}\tilde{\tau}_{\parallel}\right|%
 _{\mathrm{bare}%
}  &  =\Big(  \frac{\zeta}{2}+\frac{\zeta_{\mu}}{2}-\zeta_{\tau}\Big)
\tilde{\tau}_{\parallel}-\zeta_{y}\tau_{\perp}\,, \nonumber \\
\left.  \kappa\partial_{\kappa}\tau_{\perp}\right|  _{\mathrm{bare}}  &
=\zeta\tau_{\perp}\,.
\end{eqnarray}

One can see quite easily that $u_{\bot}^{\ast}=0$ is an invariant subspace of
the RG, since the coupling $u_{\Vert}$ alone cannot generate any corrections
to $u_{\bot}^{\ast}$. Assuming that the stable fixed point, at higher orders
of perturbation theory, remains characterized by $w^{\ast}\neq0$, $u_{\Vert
}^{\ast}\neq0$, while $u_{\bot}^{\ast}=0$, we can establish the (exact)
relations
\begin{equation}
\zeta^{\ast}=0\,,\qquad\frac{7}{4}\zeta_{\mu}^{\ast}-2\zeta_{\tau}^{\ast
}=\varepsilon\,,\qquad\zeta_{\times}^{\ast}=\frac{1}{2}\zeta_{\mu}^{\ast}\,
\label{exakt_iii}%
\end{equation}
which will lead to exponent identities, as we shall presently see. At the
fixed point, the RGE takes the simplified form
\begin{equation}
\Big[  \kappa\frac{\partial}{\partial\kappa}-\zeta_{\mu}^{\ast}\frac{\partial
}{\partial\mu_{\parallel}}+\Big(  \bigl(\frac{1}{2}\zeta_{\mu}^{\ast}-\zeta_{\tau
}^{\ast}\bigr)  \tilde{\tau}_{\parallel}-\zeta_{y}^{\ast}\tau_{\perp}\Big)
\frac{\partial}{\partial\tilde{\tau}_{\parallel}}\Big]  G_{N,\tilde{N}}=0\,.
\label{RGE_iii}%
\end{equation}
The two critical control parameters $(\tilde{\tau}_{\parallel},\tau_{\perp})$
are diagonalized by introducing the new independent variables $(\sigma
_{\parallel},\tau_{\perp})$, defined by
\begin{equation}
\sigma_{\parallel}=\tilde{\tau}_{\parallel}+\frac{2\zeta_{y}^{\ast}}%
{2\zeta_{\tau}^{\ast}-\zeta_{\mu}^{\ast}}\tau_{\perp}\,. \label{mixing}%
\end{equation}
This new critical control parameter $\sigma_{\parallel}$ takes over the role
of $\tau_{\parallel}$ in the mean-field approximation. The solution of the RGE
(\ref{RGE_iii}) is given by
\begin{eqnarray}
G_{N,\tilde{N}}(\{\mathbf{r},t\};&\tau_{\Vert}&,\tau_{\bot};\mu_{\parallel
},w^{\ast};u_{\parallel}^{\ast},u_{\perp}^{\ast};\gamma,\kappa
)= \nonumber \\
&=& F_{N,\tilde{N}}(\{\mathbf{r},t\};l^{\zeta_{\mu}^{\ast}/2-\zeta_{\tau}^{\ast
}}\sigma_{\parallel},\tau_{\bot};l^{-\zeta_{\mu}^{\ast}}\mu_{\parallel}%
;\gamma,l\kappa)\,,
\end{eqnarray}
where $l$ is an arbitrary scale, and $F_{N,\tilde{N}}$ results from
$G_{N,\tilde{N}}$ after the variable transformation to $(\sigma_{\parallel
},\tau_{\perp})$. Combining this equation with the simple parallel and
transverse scale invariances, we arrive at the scaling form
\label{G-scal_iii}%
\begin{eqnarray}
G_{N,\tilde{N}}(\{\mathbf{r},t\};\tau_{\Vert},\tau_{\bot})  &  =l^{\delta
_{N,\tilde{N}}}F_{N,\tilde{N}}(\{l^{1+\Delta}x_{\parallel},l\mathbf{r}_{\perp
},l^{z}t\};l^{-1/\nu_{\parallel}}\sigma_{\parallel},l^{-1/\nu_{\perp}}%
\tau_{\bot})\,,\label{G_iii}
\label{delta_iii}%
\end{eqnarray}
where $\delta_{N,\tilde{N}}$ is again given by Eq.~(\ref{deltaNN}) 
while the other exponents take the values 
\begin{eqnarray}
\Delta &  =\frac{\zeta_{\mu}^{\ast}}{4}\,,\qquad\eta=\zeta^{\ast}%
=0\,,\qquad z=4-\eta = 4\,,\nonumber \\
1/\nu_{\parallel}  &  =2-\frac{\zeta_{\mu}^{\ast}}{2}+\zeta_{\tau}^{\ast
}\,,\qquad \qquad \,\,\, 1/\nu_{\perp}=2\,.
\end{eqnarray}
The result to all orders, Eq.~(\ref{exakt_iii}), leads to an exact relation
between $\Delta$ and $\nu_{\parallel}$:
\begin{equation}
\nu_{\Vert}=\frac{2}{d-2+3\Delta}\,. \label{nu_par_iii}%
\end{equation}
Once again, only a single exponent must be computed \textit{explicitly} within
the $\varepsilon$-expansion, e.g., $\Delta$. Our one-loop calculation yields
\begin{equation}
\Delta=\frac{23+6\sqrt{15}}{11}\varepsilon+O(\varepsilon^{2})\,,
\end{equation}
and for the mixing of the critical control parameters
\begin{equation}
\sigma_{\parallel}=\frac{\tau_{\parallel}}{\sqrt{\mu_{\parallel}}}+\Big(
\frac{17}{3\bigl(  10+\sqrt{15}\bigr)  }+O(\varepsilon)\Big)  \tau_{\perp}\,.
\end{equation}
\qquad\qquad

Let us briefly return to the assumption that the stable fixed point is
characterized by $u_{\perp}^{\ast}=0$. Thanks to Eq.~(\ref{beta_iii}) and the
associated linear stability matrix, we can determine the stability criterion,
i.e., the correction-to-scaling exponent of $u_{\perp}$, to all orders in
perturbation theory. The condition for $u_{\perp}^{\ast}=0$ to be stable is
$4\varepsilon<3\zeta_{\mu}^{\ast}$ which generates bounds for the critical
exponents $\Delta$ and $\nu_{\parallel}$, namely,%
\begin{equation}
\nu_{\parallel}<\frac{1}{2}\,,\qquad\Delta>\frac{6-d}{3}\,. \label{Bdg-1}%
\end{equation}
If the stability condition $4\varepsilon<3\zeta_{\mu}^{\ast}$ or equivalently,
the second bound of Eq.~(\ref{Bdg-1}), is violated at some dimension $d$, the
fixed point structure of the theory changes fundamentally. Now $u_{\perp
}^{\ast}\neq0$ is stable, and Eq.~(\ref{beta_iii}) gives us the (exact)
relations
\begin{equation}
4\varepsilon-3\zeta_{\mu}^{\ast}=9\zeta^{\ast}>0\,,\qquad\zeta_{\times
}^{\ast}=\frac{1}{2}\left(  \zeta^{\ast}+\zeta_{\mu}^{\ast}\right)
\end{equation}
instead of Eqs.~(\ref{exakt_iii}). We lose one condition, and this leads to
two independent critical exponents, as opposed to a single one. Following the
same analysis as before, we arrive at the relations
\label{exp_iii-unst}%
\begin{eqnarray}
\Delta  =\frac{6-d}{3}-\eta\,,\quad\eta>0\,, \quad
1/\nu_{\perp}   =2-\eta\,. \label{exp_iii-unst(2)}%
\end{eqnarray}
The scaling behavior of the Green functions still follows Eq.
(\ref{G_iii}), but $\nu_{\Vert}$ can no longer be related to $\Delta$.


\section{The roughness exponents}

The roughness exponents of the surface are easily identified, once the scaling
properties of the underlying field theory are known. For isotropic theories,
they can be read off directly from the height-height correlation function,
\begin{eqnarray}
G_{2,0}(\mathbf{r}-\mathbf{r}^{\prime},t-t^{\prime})\equiv C(\mathbf{r}%
-\mathbf{r}^{\prime},t-t^{\prime})\equiv\left\langle h(\mathbf{r}%
,t)h(\mathbf{r}^{\prime},t^{\prime})\right\rangle \nonumber
\end{eqnarray}
if its asymptotic scaling behavior of can be
written in the form
\begin{equation}
C(\mathbf{r},t)=\left|  \mathbf{r}\right|  ^{2\chi}c(t/\left|  \mathbf{r}%
\right|  ^{z}) \label{isotropic_realspace_roughness}%
\end{equation}
Here, $\chi$ denotes the roughness exponent and $z$ the dynamic exponent of
the surface while $c$ is a universal scaling function. Similarly, in Fourier
space, the behavior of $C(\mathbf{r},t)$ translates into
\begin{equation}
\widetilde{C}(\mathbf{q},t)=\left|  \mathbf{q}\right|  ^{-(d+2\chi)}%
\widetilde{c}\left(  t\left|  \mathbf{q}\right|  ^{z}\right)
\label{isotropic_kspace_roughness}%
\end{equation}
with the same roughness exponent $\chi$. In our case, the situation is
slightly more subtle \cite{SPJ_PRB}, due to the presence of 
strong anisotropy \cite{SZ:1995}.
First of all, surface fluctuations along the parallel and the transverse
directions in real space need not be controlled by the same indices, leading
us to define two exponents, $\chi_{\bot}$ and $\chi_{\Vert}$, via
\begin{eqnarray}
C(x_{\Vert}  =0,\mathbf{r}_{\bot};t) &\equiv& \left|  \mathbf{r}_{\bot}\right|
^{2\chi_{\bot}}c_{\bot}(t/\left|  \mathbf{r}_{\perp}\right|  ^{z}) \nonumber \\
C(x_{\Vert},\mathbf{r}_{\bot}   =\mathbf{0};t) &\equiv& \left|  x_{\Vert
}\right|  ^{2\chi_{\Vert}}c_{_{\Vert}}\bigl(  t/x_{\Vert}^{z/(1+\Delta
)}\bigr)
\end{eqnarray}
A similar situation should be expected in Fourier space, prompting us to
define two additional exponents, $\tilde{\chi}_{\bot}$ and $\tilde{\chi
}_{\Vert}$, via
\label{aniso_MS_scaling}%
\begin{eqnarray}
\widetilde{C}(0,\mathbf{q}_{\bot},t)  &  \equiv\left|  \mathbf{q}_{\bot
}\right|  ^{-(d+2\tilde{\chi}_{\bot})} \, \widetilde{c}_{\bot}\left(
t\left|  \mathbf{q}_{\bot}\right|  ^{z}\right) \nonumber \\
\widetilde{C}(q_{\Vert},\mathbf{0},t)  &  \equiv q_{\Vert}^{-(d+2\tilde{\chi
}_{\Vert})} \, \widetilde{c}_{_{\Vert}}\bigl(  tq_{\Vert}^{z/(1+\Delta
)}\bigr)
\end{eqnarray}
Of course, all of these expressions are only meaningful if the four scaling
functions $c_{\bot}$, $c_{_{\Vert}}$, $\widetilde{c}_{\bot}$ and
$\widetilde{c}_{\Vert}$ approach finite and non-zero constants when their
arguments vanish. Generically, the four roughness exponents take different
numerical values; however, they can all be expressed in terms of the exponents
$\Delta$ and $\eta$, according to the identities
\begin{eqnarray}
\chi_{\bot}=\frac{1}{2}\bigl[  4-\left(  d+\Delta\right)  -\eta\bigr]  \quad
\quad 
\chi_{\Vert}=\frac{1}{2}\bigl[  4-\left(  d+\Delta\right)
-\eta\bigr]/ (1+\Delta) \label{Chi_RS} \nonumber \\
\tilde{\chi}_{\bot}=\frac{1}{2}\bigl[  4-d-\eta\bigr]  \qquad \qquad \,\,\,
\tilde
{\chi}_{\Vert}=\frac{1}{2}\bigl[  4-\eta\bigr]/ (1+\Delta)
- \frac{d}{2} 
\label{Chi_MS}%
\end{eqnarray}
The key observation is that $\tilde{\chi}_{\bot}=\tilde{\chi}_{\Vert}=\chi_{\bot
}=\chi_{\Vert}$ only if the anisotropy exponent $\Delta$ vanishes.

In the following, we explicitly compute the roughness exponents for the two
cases (i) and (iii). For case (ii), the renormalization group gives us no such
information, since that would require the existence of an infrared stable
fixed point. If, in fact, the presence of a first order transition line were
to be confirmed eventually, the whole concept of scaling exponents would be
inapplicable here.

\noindent {\it Case} (i) For this theory, characterized by positive
$\tau_{\parallel}$ while $\tau_{\perp}$ vanishes, we found that all
exponents could be expressed in terms of 
$\eta=-\frac{2}{3}\varepsilon+O(\varepsilon^{2})$. Writing all four roughness
exponents in terms of this single index, we arrive at expressions 
which are exact to all orders in $\varepsilon=3-d$:
\begin{eqnarray}
\chi_{\bot}  &=& 1-d/3\,, \qquad  \chi_{\Vert}  =\frac{1-d/3}{3-d/3-\eta
}\,,\label{Chi_RS2} \nonumber \\
\tilde{\chi}_{\bot}    &=& (4-d-\eta)/2\,, \quad \tilde{\chi}_{\Vert} 
=  \frac{4-\eta}{2(3-d/3-\eta)} - \frac{d}{2}\,.
\label{Chi_MS2}%
\end{eqnarray}
The mean-field values are easily recovered by setting $\varepsilon=0$. The
physically most interesting case corresponds to a surface grown on a
two-dimensional substrate, i.e., $d=2$ and $\varepsilon=1$. For this
situation, one obtains $\chi_{\bot}=1/3$ and $\chi_{\Vert}=1/9+O(\varepsilon
^{2})$ while $\tilde{\chi}_{\Vert}=-2/9+O(\varepsilon^{2})$ and $\tilde{\chi
}_{\bot}=4/3+O(\varepsilon^{2})$. Remarkably, the exponent $\chi_{\bot}=1/3$
is actually \textit{exact}, at least to all orders in perturbation theory.

\noindent {\it Case} (iii): At the multicritical point one has 
$\eta=0$, $z=4$, and a nontrivial $\Delta$. Hence, one obtains

\begin{eqnarray}
\chi_{\bot}= \bigl[  4-\left(  d+\Delta\right)\bigr]/2  \quad
\quad 
\chi_{\Vert}=\frac{4-\left(  d+\Delta\right)}{2(1+\Delta)} 
\label{Chi_RS3} \nonumber \\
\tilde{\chi}_{\bot}=\bigl[  4-d\bigr]/2  \qquad \qquad \,\,\,
\tilde
{\chi}_{\Vert}= \frac{2}{(1+\Delta)}
- \frac{d}{2} 
\label{Chi_MS3}%
\end{eqnarray}

All of them are negative near the upper critical dimension $d_{c}=6$. In order
to access the physical ($d=2$) situation, one has to set $\varepsilon=4$ here
which gives a huge anisotropy exponent, $\Delta\simeq16.814$, if one naively
uses the one-loop results. While roughness exponents can in principle be calculated,
it does not appear meaningful to compare them to experimental data. Yet, we
emphasize again that the exponents calculated here do not agree 
with \cite{MarsiliETAL:1996}.


\section{Conclusions}

To summarize, we have analyzed the effect of strong anisotropies on the
universal behavior of a surface grown under ideal MBE-type conditions. The
anisotropy is generated by an incident particle beam which is tilted away from
the normal. This arrangement manifestly breaks rotational invariance in the
substrate plane. As a
consequence, the effective surface tension becomes anisotropic, being
characterized by two parameters, $\tau_{\parallel}$ and $\tau_{\perp}$, as
opposed to a single one in the isotropic case. If both are positive, the
surface is described by the Edwards-Wilkinson model. However, as temperature
or other growth parameters are varied, the two control parameters need not
vanish simultaneously, leading to different dynamic and spatial instabilities
and ordered structures. Clearly, there are three distinct scenarios:\ (i)
$\tau_{\parallel}>0$ while $\tau_{\perp}$ goes to zero, (ii)$\;\tau
_{\parallel}\rightarrow0$ while $\tau_{\perp}$ remains positive, and (iii)
both parameters vanish. Each of these three cases is described by its own
characteristic field theory with distinct upper critical dimensions:\ $d_{c}%
=3$, $7/2$, and $6$ for cases (i), (ii), and (iii), respectively. To display
scale-invariant behavior in the long-time, large-distance limit, a field
theory must possess one, or more, infrared stable fixed points. Our one-loop
analysis reveals that only cases (i) and (iii) have this property. In
contrast, case (ii) may correspond to a line of first-order phase transition
whose characteristics lie outside the scope of our RG techniques. For the two
scale-invariant theories, we find two distinct, novel surface universality
classes, and derive the associated scaling behavior of the 
height-height correlation function. An intriguing feature of both universality
classes is the emergence of a strong anisotropy exponent, which reflects
different scaling behavior in the parallel and transverse directions. As an
immediate consequence, we need to introduce four different roughness
exponents, to characterize surface fluctuations in real and momentum space,
observed along the parallel vs the transverse directions. While all four are
related by exponent identities, care must nevertheless be exercised when
analyzing experimental data.

We conclude with a few comments on the phase diagram. The first case clearly
corresponds to a second order line, parameterized by $\tau_{\parallel}$.
Assuming that the first order nature of the second case, parameterized by
$\tau_{\perp}$, can be confirmed, the third case $\tau_{\parallel}=\tau
_{\perp}=0$ would turn out to be a critical endpoint. Since it requires the
tuning of two parameters, it may be quite difficult to access in a typical
experiment. In contrast, either the second or the first order lines should be
more easily observable. From an RG perspective, even if the $\tau_{\parallel
}\partial_{\parallel}^{2}h$ term is absent initially, it is immediately
generated under RG transformations, leading to a non-zero critical value of
$\tau_{\parallel}$; this is not the case for the $\tau_{\perp}\nabla_{\perp
}^{2}h$ contribution. For this reason, we believe that the most physically
relevant theory (beyond Edwards-Wilkinson behavior)\ is the one with
$\tau_{\parallel}>0$ and $\tau_{\perp}=0$. Our key results for this model are
the roughness exponents for real-space surface scans. For a two-dimensional
surface embedded in a three-dimensional space, we find $\chi_{\bot}=1/3$ for
scans along the transverse direction, and $\chi_{\Vert}=1/9+O(\varepsilon
^{2})$ for parallel scans. There appears to be some experimental evidence for
these exponents \cite{SPJ_PRB} but more data are needed before this issue
can be settled.

\vspace{0.2cm}
We thank U.C. T\"{a}uber, R.K.P. Zia, J. Krug, A. Hartmann,
and E. Yewande for helpful discussions. This work is partially supported by
the NSF through DMR-0308548 and DMR-0414122. GP acknowledges the
Alexander von Humboldt foundation for their support.

\appendix                        

\section*{Appendix}

In the following, some technical details of the above calculations are
presented. We prefer the time-momentum representation of the propagators and
correlators, since it displays the causal structure of the Feynman diagrams
most directly. In this representation, each $\Gamma_{1,1}$-\ (or self
energy)\ diagram has a single $\tilde{h}$-leg at the outgoing vertex with the
largest time argument, and a single $h$-leg at the incoming vertex with the
smallest time argument. To find the UV divergent $\omega=0$ contribution, we
just integrate over the time difference of these two vertices.

\section{Case (i): $\tau_{\perp}\to0$ and $\tau_{\parallel}\ne0$}

\label{ParFinitePerpVanish}

To characterize the theory fully, it suffices to calculate the vertex
function $\Gamma_{1,1}$. The only one loop diagram 
which must be evaluated is
\begin{equation}
\Gamma_{1,1}(\mathbf{q},\omega)=G_{0}(-\mathbf{q},-\omega)^{-1}-
\parbox[c][1cm]{1.6cm}{\psarc (0.8,0){0.55}{0}{360}%
\psarc{<-}(0.8,0){0.55}{270}{0}%
\psline{<-}(0,0)(0.25,0) \psline (1.35,0)(1.6,0)}
+ \dots
\label{one_loop_Gamma11_tauperp0}%
\end{equation}
using the diagrammatic building blocks associated with 
Eqs.~(\ref{def_bare_props}) and (\ref{full_vertices}). This graph
represents the one loop contribution to the self-energy. After integrating
over the time difference of the two vertices, it takes the form
\begin{equation}
\parbox[c][1 cm]{2cm}{\psarc (0.8,0){0.55}{0}{360}\psarc{<-} (0.8,0){0.55}%
{270}{0}\psline{<-}(0,0)(0.25,0)\psline (1.35,0)(1.6,0)}=-\gamma\lambda
_{\bot}^{2}\int\frac{d^{d}p}{(2\pi)^{d}}\frac{\bigl[ -(\mathbf{q}_{\bot}%
^{2})^{2}p_{\Vert}^{2}+2\mathbf{q}_{\bot}^{2}(\mathbf{q}_{\bot}\cdot
\mathbf{p}_{\bot})p_{\Vert}^{2}+q_{\Vert}^{2}(\mathbf{p}_{\bot}^{2})^{2}%
\bigr] }{\Gamma(\mathbf{p})\left[  \Gamma(\mathbf{p})+\Gamma
(\mathbf{q}-\mathbf{p})\right]  \,.}\label{integral1_case_tperp0}%
\end{equation}
Several finite terms have been dropped, following a dimensional
regularization and minimal subtraction scheme. The function
$\Gamma(\mathbf{p})$ contains, up to a factor $\gamma$, the full momentum
dependence of the inverse propagator
\begin{equation}
\Gamma(\mathbf{p})=\mathbf{p}_{\bot}^{2}(\mathbf{p}_{\bot}^{2}+\tau_{\bot
})+p_{\Vert}^{2}\tau_{\Vert}\ .\label{prop_i}%
\end{equation}

Dimensional analysis immediately suggests that the terms in the bracket of
Eq.~(\ref{integral1_case_tperp0}) diverge logarithmically, except possibly the
term $2\mathbf{q}_{\bot}^{2}(\mathbf{q}_{\bot}\cdot\mathbf{p}_{\bot})p_{\Vert
}^{2}$, which is, at least naively, linearly divergent. The denominator
in Eq.~(\ref{integral1_case_tperp0}) is now expanded in small external
momenta, 
and multiplied with the numerator, keeping only
divergent contributions. 
After some straightforward algebra, we arrive at
\begin{equation}
\mathrm{(\ref{integral1_case_tperp0})}=-\gamma\lambda_{\bot}^{2}\left[
-\frac{1}{2}(\mathbf{q}_{\perp}^{2})^{2}I(2,0,2)+\frac{1}{2}q_{\Vert}^{2}%
I(0,4,2)+\frac{2}{d-1}(\mathbf{q}_{\perp}^{2})^{2}I(2,4,3)\right]
\label{int}
\end{equation}
The integrals $I(\alpha,\beta,\gamma)$ are defined as
\begin{equation}
I(\alpha,\beta,\gamma)\equiv\int\frac{d^{d}p}{(2\pi)^{d}}\frac{p_{\Vert
}^{\alpha}(\mathbf{p}_{\bot}^{2})^{\beta/2}}{\Gamma(p)^{\gamma}}\propto
\Lambda^{d+1+2\alpha+\beta-4\gamma}\label{integral_iii}%
\end{equation}
with the proportionality indicating the degree of divergence based on
simple power counting in $d=d_{c}=3$. 
Clearly, all integrals appearing in Eq.~(\ref{int}) are
only logarithmically divergent; the (naive) linear divergence
vanishes by symmetry.
Using the effective coupling $u$ introduced in Eq.~(\ref{def_u_i}), the
integrals are
\begin{equation}
\lambda_{\bot}^{2}I(2,0,2)=\frac{u^{2}}{2\varepsilon}\,,\quad\lambda_{\bot
}^{2}I(0,4,2)=\frac{u^{2}}{2\varepsilon}\tau_{\Vert}\,,\quad\lambda_{\bot}%
^{2}I(2,4,3)=\frac{u^{2}}{8\varepsilon}\,,
\end{equation}
and one finally arrives at Eq.~(\ref{Gamma11_i}).

\section{Case (ii): $\tau_{\parallel}\to0$ and $\tau_{\perp}\ne0$}

\label{ParVanishPerpFinite} The diagrammatic representation of $\Gamma_{1,1}$,
to one loop, is again given by Eq.~(\ref{one_loop_Gamma11_tauperp0}). However,
the relevant interaction vertex here is $\lambda_{\parallel}$, so that the one
loop integral reads
\begin{equation}
\gamma\lambda_{\parallel}^{2}q_{\parallel}^{2}\int\frac{d^{d}p}{(2\pi)^{d}%
}p_{\parallel}^{2}(q_{\parallel}-p_{\parallel})^{2}\frac{1}{\Gamma
(\mathbf{p})\left[  \Gamma(\mathbf{p})+\Gamma(\mathbf{q}-\mathbf{p})\right]
}\label{Gamma11_integral1_case_tpar0}%
\end{equation}
and the momentum dependence of the inverse bare propagator is controlled by
\begin{equation}
\Gamma(\mathbf{p})=p_{\Vert}^{2}(\mu_{\parallel}p_{\Vert}^{2}+\tau_{\Vert
})+\tau_{\bot}\mathbf{p}_{\bot}^{2}\ .\label{prop_ii}%
\end{equation}
As in the previous section, the integral (\ref{Gamma11_integral1_case_tpar0})
cannot be performed in closed form. Again,
we expand the denominator of Eq.~(\ref{Gamma11_integral1_case_tpar0}) in
powers of $q_{\parallel}$ and $\mathbf{q}_{\perp}$, noting that higher orders
become more and more convergent. Keeping only divergent terms, the singular
part of $\Gamma_{1,1}$ takes the form  \
\begin{eqnarray}
\Gamma_{1,1} &  =i\omega+\gamma\left[  q_{\Vert}^{2}(\mu_{\parallel}q_{\Vert
}^{2}+\tau_{\Vert})+\tau_{\bot}\mathbf{q}_{\bot}^{2}\right]  -\gamma
\lambda_{\Vert}^{2}q_{\Vert}^{2}\int\frac{d^{d}p}{(2\pi)^{d}}\frac{p_{\Vert
}^{4}}{2\Gamma(\mathbf{p})^{2}}\\
&  -\gamma\lambda_{\parallel}^{2}q_{\parallel}^{4}\int\frac{d^{d}p}{(2\pi
)^{d}}\left\{  \frac{p_{\Vert}^{2}}{2\Gamma(\mathbf{p})^{2}}-\frac{7\mu
_{\parallel}p_{\Vert}^{6}}{2\Gamma(\mathbf{p})^{3}}+\frac{4\mu_{\parallel}%
^{2}p_{\Vert}^{10}}{2\Gamma(\mathbf{p})^{4}}+\dots\right\}  +O(\lambda
_{\parallel}^{4})\ .\nonumber
\end{eqnarray}
The leading singularity, i.e., the first integral above, is found to be
quadratically divergent. It contributes to the renormalization of
$\tau_{\parallel}$ and can therefore be controlled by an \textit{additive}
renormalization. We shift the critical parameter by an amount $\mathring{\tau
}_{\parallel,c}$, defined through the vanishing of the parallel
susceptibility, Eq.~(\ref{susc}). Evaluating $\chi_{\Vert}$ 
to first order yields 
\begin{equation}
\chi_{\Vert}^{-1}=\tau_{\Vert}-\lambda_{\Vert}^{2}\int\frac{d^{d}p}{(2\pi
)^{d}}\frac{p_{\Vert}^{4}}{2\Gamma^{2}(\mathbf{p})}+O(\lambda_{\parallel}%
^{4})\ .
\end{equation}
The general renormalization scheme, Eq.~(\ref{gen_ren_scheme}), 
then allows us to determine
\begin{equation}
\mathring{\tau}_{\parallel,c}=\left.  \lambda_{\Vert}^{2}\int\frac{d^{d}%
p}{(2\pi)^{d}}\frac{p_{\Vert}^{4}}{2\Gamma^{2}(p)}\right|  _{\tau_{\Vert}%
=0}+O(\lambda_{\Vert}^{4})\,.
\end{equation}
in one loop. Reparameterizing $\Gamma_{1,1}$ in terms of the shifted
 $\tau_{\Vert}$ subsequently removes the quadratic divergence and leaves 
us with logarithmic divergences only. 
Hence, the vertex function reads, to first order,
\begin{eqnarray*}
\Gamma_{1,1} &  =i\omega+\gamma\left(  q_{\Vert}^{2}(\mu_{\parallel}q_{\Vert
}^{2}+\tau_{\Vert})+\tau_{\bot}\mathbf{q}_{\bot}^{2}\right)
+\gamma\lambda_{\Vert}^{2}q_{\Vert}^{2}\hat{\tau}_{\Vert}J(6,3)\\
&  -\frac{1}{2}\gamma\lambda_{\Vert}^{2}q_{\Vert}^{4}\left[  J(2,2)-7\mu
_{\parallel}J(6,3)+4\mu_{\parallel}^{2}J(10,4)+...\right]
\end{eqnarray*}
with the integrals $J(\alpha,\gamma)$ defined by
\begin{equation}
J(\alpha,\gamma)\equiv\int\frac{d^{d}p}{(2\pi)^{d}}\frac{p_{\Vert}^{\alpha}%
}{\Gamma(\mathbf{p})^{\gamma}}\propto\Lambda^{2d-1+\alpha-4\gamma
}\label{integral_ii}%
\end{equation}
Again, the proportionality indicates the degree of divergence based on
$d=d_{c}=7/2$. Using the effective coupling $u$ introduced in
Eq.~(\ref{effective_coupl_case_tpar0}), the integrals can be evaluated:
\begin{equation}
\lambda_{\Vert}^{2}J(2,0,2)=\mu_{\parallel}\frac{4u^{2}}{3\varepsilon}%
\,,\quad\lambda_{\Vert}^{2}J(6,0,3)=\frac{u^{2}}{2\varepsilon}\,,\quad
\lambda_{\Vert}^{2}J(10,0,4)=\mu_{\parallel}^{-1}\frac{7u^{2}}{24\varepsilon
}\,,
\end{equation}
which finally results in Eq.~(\ref{eq:Gamma11_ii}).

\section{Case (iii): $\tau_{\parallel}\to0$ and $\tau_{\perp}\to0$}

\label{ParVanishPerpVanish} The calculation of the self energy follows the
same standard methods as above. However, the technical details become more
involved, for two reasons. First, the full momentum dependence of the inverse
propagator,
\begin{equation}
\Gamma(\mathbf{p})=\mathbf{p}_{\perp}^{2}(\mathbf{p}_{\perp}^{2}+\tau_{\perp
})+2\mathbf{p}_{\perp}^{2}p_{\parallel}^{2}\mu_{\times}+p_{\parallel}^{2}%
(\mu_{\parallel}p_{\parallel}^{2}+\tau_{\parallel})\label{prop_iii}%
\end{equation}
comes into play, with the two critical parameters, $\tau_{\perp}$ and
$\tau_{\parallel}$, serving as infrared cutoffs. Second, both nonlinearities,
parametrized by $\lambda_{\parallel}$ and $\lambda_{\perp}$, are now marginal
at the upper critical dimension $d_{c}=6$.

As before, we start with the diagrammatic representation, Eq.
(\ref{one_loop_Gamma11_tauperp0}). Due to the presence of the full vertex, Eq.
(\ref{full_vertices}), the one-loop integral is given by
\[
-\gamma\int\frac{d^{d}p}{(2\pi)^{d}}\frac{V(\left\{  \mathbf{q},\mathbf{p}%
\right\}  )}{\Gamma(\mathbf{p})\left[  \Gamma(\mathbf{p})+\Gamma
(\mathbf{q}-\mathbf{p})\right]  }%
\]
where $V(\left\{  \mathbf{q},\mathbf{p}\right\}  )$ summarizes the momenta
contributed by the vertices:%
\begin{eqnarray*}
V(\left\{  \mathbf{q},\mathbf{p}\right\}  )   \equiv &-& \lambda_{\Vert}%
^{2}q_{\Vert}^{2}p_{\Vert}^{2}\left(  q_{\Vert}-p_{\Vert}\right)  ^{2}\\
&+& \lambda_{\bot}^{2}\left\{  \left[  2(\mathbf{q}_{\bot}\cdot\mathbf{p}%
_{\bot})p_{\Vert}-\mathbf{q}_{\bot}^{2}p_{\Vert}-q_{\Vert}\mathbf{p}_{\bot
}^{2}\right]  \left[  \mathbf{q}_{\bot}^{2}p_{\Vert}-q_{\Vert}\mathbf{p}%
_{\bot}^{2}\right]  \right\}  \\
&+& \lambda_{\Vert}\lambda_{\bot}q_{\Vert}\left(  q_{\Vert}-p_{\Vert}\right)
p_{\Vert}\left[  2(\mathbf{q}_{\bot}\cdot\mathbf{p}_{\bot})p_{\Vert
}-2\mathbf{q}_{\bot}^{2}p_{\Vert}\right]
\end{eqnarray*}
As for case (ii), a shift of the critical parameter $\tau_{\parallel}$ absorbs
a quadratic divergence in $\Gamma_{1,1}$. Demanding that the inverse
(parallel) susceptibility, Eq.~(\ref{susc}) vanishes at the true critical
point, one finds in one loop order
\begin{eqnarray}
\mathring{\tau}_{\Vert,c} = \lambda_{\Vert}^{2}\left.  \int\frac{d^{d}p}%
{(2\pi)^{d}}\frac{p_{\Vert}^{4}}{2\Gamma(p)^{2}}\right|  _{\tau_{\Vert}%
=\tau_{\bot}=0} 
- \lambda_{\bot}^{2}\left.  \int\frac{d^{d}p}{(2\pi)^{d}%
}\frac{(\mathbf{p}_{\bot}^{2})^{2}}{2\Gamma(p)^{2}}\right|  _{\tau_{\Vert
}=\tau_{\bot}=0} \label{shift_iii}%
\end{eqnarray}
Again, we retain only logarithmic divergences in $\Gamma_{1,1}$ if we
reparameterize it in terms of a shifted $\tau_{\Vert} \rightarrow 
\tau_{\parallel}%
-\mathring{\tau}_{\parallel,c}$. Expanding the denominator to $O(\mathbf{q}%
^{2})$ results in a sum of integrals of the form
\begin{equation}
I(\alpha,\beta,\gamma)\equiv\int\frac{d^{d}p}{(2\pi)^{d}}\frac{p_{\Vert
}^{\alpha}(\mathbf{p}_{\bot}^{2})^{\beta/2}}{\Gamma(p)^{\gamma}}\propto
\Lambda^{d+\alpha+\beta-4\gamma}\label{key_integral_iii}%
\end{equation}
Formally, these look like Eq.~(\ref{integral_iii}), but it is essential to
realize that $\Gamma(\mathbf{p})$ takes the more complicated form
(\ref{prop_iii}). To evaluate the $I(\alpha,\beta,\gamma)$, it is convenient
to introduce a new integration variable, $x\equiv q_{\parallel}/\left|
\mathbf{q}_{\perp}\right|  $, to replace $q_{\parallel}$. This has
the significant advantage that all $\varepsilon$-poles are already explicitly
displayed, once the $\mathbf{q}_{\perp}$-integration has been performed. The
remaining integrals over $x$ can be evaluated at $\varepsilon=0$ and are
perfectly straightforward. They take the form
\begin{equation}
K_{n,\sigma}(w)=\int_{-\infty}^{\infty}dx\,\frac{x^{2\sigma}}{\bigl(
1+2wx^{2}+x^{4}\bigr) }%
\end{equation}
with $(n,\sigma)\in\{(1,0),(2,0),(3,0),(4,0),(2,1),(3,1),(4,1)\}$. Clearly,
these $K$'s lead to $w$-dependent coefficients for the different contributions
to $\Gamma_{1,1}$. The result is summarized in Eq.~(\ref{Gamma11_iii})
with the $w$-dependent functions
\label{parameter_integrals_caseIII}%
\begin{eqnarray}
A_{1} &  =2\left(  1+w\right)  \,,\nonumber\\
B_{1} &  =-3\,,\quad B_{2}=(5+2w)\,,\quad B_{3}=3\left(  7+2w(5+2w)\right)
\,,\nonumber\\
C_{1} &  =-\left(  3+w\right)  \,,\quad C_{2}=\left(  5+9w+10w^{2}%
+4w^{3}\right)  \,,\label{w_fcts}\\
D_{1} &  =-\frac{1}{5}\left(  7+5w\right)  \,,\quad D_{2}=\frac{1}{5}\left(
49+65w+26w^{2}\right)  \,,\quad D_{3}=\frac{1}{5}\left(  10+2w\right)
\,.\nonumber
\end{eqnarray}
Tallying up all the contributions to $\Gamma_{1,1}$, and removing 
the divergences by renormalization, one arrives at the $Z$-factors
listed in Eq.~(\ref{Z_iii}). 

\vspace{0.5cm}

\end{document}